\documentclass[final,1p,times,twocolumn]{elsarticle}

\journal{Journal of Magnetism and Magnetic Materials}

\usepackage{amssymb}
\usepackage{hyperref}
\usepackage{graphicx}
\usepackage{amsmath,amssymb,amsfonts}

\begin{document}

\begin{frontmatter}

\title{Thermodynamic and transport properties of single crystalline RCo$_{2}$Ge$_{2}$ (R = Y, La-Nd, Sm-Tm)}

\author[label1,label2]{Tai Kong} \author[label3]{Charles E. Cunningham} \author[label1]{Valentin Taufour} \author[label1,label2]{Sergey L. Bud'ko} \author[label2]{Malinda L. C. Buffon} \author[label1,label2]{Xiao Lin} \author[label3]{Heather Emmons} \author[label1,label2]{Paul C. Canfield}

\address[label1]{Department of Physics and Astronomy, Iowa State University, Ames, Iowa 50011, U.S.A.}

\address[label2]{Ames Laboratory, Iowa State University, Ames, Iowa 50011, U.S.A.}

\address[label3]{Department of Physics, Grinnell College, Grinnell, Iowa 50112, U.S.A.}

\begin{abstract}

Single crystals of RCo$_{2}$Ge$_{2}$ (R = Y, La-Nd, Sm-Tm) were grown using a self-flux method and were characterized by room-temperature powder x-ray diffraction; anisotropic, temperature and field dependent magnetization; temperature and field dependent, in-plane resistivity; and specific heat measurements. In this series, the majority of the moment-bearing members order antiferromagnetically; YCo$_{2}$Ge$_{2}$ and LaCo$_{2}$Ge$_{2}$ are non-moment-bearing. Ce is trivalent in CeCo$_{2}$Ge$_{2}$ at high temperatures, and exhibits an enhanced electronic specific heat coefficient due to Kondo effect at low temperatures. In addition, CeCo$_{2}$Ge$_{2}$ shows two low-temperature anomalies in temperature-dependent magnetization and specific heat measurements. Three members (R = Tb-Ho) have multiple phase transitions above 1.8 K. Eu appears to be divalent with total angular momentum \textit{L} = 0. Both EuCo$_{2}$Ge$_{2}$ and GdCo$_{2}$Ge$_{2}$ manifest essentially isotropic paramagnetic properties consistent with \textit{J} = \textit{S} = 7/2. Clear magnetic anisotropy for rare-earth members with finite \textit{L} was observed, with ErCo$_{2}$Ge$_{2}$ and TmCo$_{2}$Ge$_{2}$ manifesting planar anisotropy and the rest members manifesting axial anisotropy. The experimentally estimated crystal electric field (CEF) parameters B$_{2}^{0}$ were calculated from the anisotropic paramagnetic $\theta_{ab}$ and $\theta_{c}$ values and follow a trend that agrees well with theoretical predictions. The ordering temperatures, T$_{N}$, as well as the polycrystalline averaged paramagnetic Curie-Weiss temperature, $\Theta_{avg}$, for the heavy rare-earth members deviate from the de Gennes scaling, as the magnitude of both are the highest for Tb, which is sometimes seen for extremely axial systems. Except for SmCo$_{2}$Ge$_{2}$, metamagnetic transitions were observed at 1.8 K for all members that ordered antiferromagnetically. 

\end{abstract}


\end{frontmatter}

\section{Introduction}

The RT$_{2}$X$_{2}$ (R = Y, La-Lu; T = transition metal; X = Si, Ge) ternary intermetallic family has been studied for decades\cite{book}. Most of RT$_{2}$X$_{2}$ compounds crystallize in the ThCr$_{2}$Si$_{2}$ tetragonal structure\cite{Ban1} (space group I4/mmm) with a single R ion site with a tetragonal point symmetry. Given that transition metals in this family, except for Mn, are non-moment-bearing, the magnetic properties of these compounds are mainly determined by rare earths' local moment, the temperature dependent single ion anisotropy as well as the long range, indirect interactions (Ruderman-Kittel-Kasuya-Yosida (RKKY) type) between local moments via conduction electrons\cite{book}. The competition between Fermi surface nesting (maxima in the generalized magnetic susceptibility)\cite{islam} and local moment anisotropy can lead to either incommensurate or commensurate magnetic propagation vectors, or sometimes multiple transitions from one to the other on cooling. Although detailed anisotropic studies of the RNi$_{2}$Ge$_{2}$ and RFe$_{2}$Ge$_{2}$ series have been made\cite{RNi,RFe}, studies of RCo$_{2}$Ge$_{2}$ compounds and their magnetic properties have primarily been done on polycrystalline samples with magnetic ordering temperature determined only down to 4.2 K\cite{mccall1,mccall2,Pinto}. So far, only members with R = Ce\cite{Cefujii,Cepaul}, Pr\cite{Prvej}, Nd\cite{Cefujii}, Eu\cite{EuHossain,EuDio}, Gd\cite{GdGood} and Tb\cite{Tbshigeoka1,Tbshigeoka,Tbwiener} had been studied in single crystal form. The systematic growth and study of single crystals can address the issue of anisotropy and possible metamagnetic transitions in this series of compounds. It also allows for comparison to the aforementioned RNi$_{2}$Ge$_{2}$ and RFe$_{2}$Ge$_{2}$ series.

According to published literature, CeCo$_{2}$Ge$_{2}$ does not undergo any magnetic transitions and is reported to exhibit Kondo screening of the 4f moment. No consensus was reached on Ce's valence state\cite{Cefujii,Cepaul,Cex,takashi}. Neutron diffraction work on PrCo$_{2}$Ge$_{2}$ indicates a magnetic ordering at $\sim$ 28 K with a sinusoidally modulated incommensurate magnetic structure along \textit{c}-axis\cite{Pr,PrHo}. In addition, at 2 K, well below its ordering temperature, two metamagnetic transitions were observed at 20 kOe and 100 kOe\cite{Prvej,Prvin}. NdCo$_{2}$Ge$_{2}$ has been reported to order antiferromagnetically at around 30 K and undergo a transition between different magnetic structures at $\sim$10 K\cite{Cefujii,NdEr,Ndcp,Ndandre}. Two metamagnetic transitions were observed at $\sim$ 30 kOe and $\sim$ 110 kOe\cite{Prvin}. Temperature-dependent susceptibility and specific heat were measured on polycrystalline SmCo$_{2}$Ge$_{2}$\cite{Sm}, which indicate a magnetic transition at around 14 K. Single crystals of EuCo$_{2}$Ge$_{2}$ were synthesized using self flux method and characterized by transport measurements. A paramagnetic to antiferromagnetic transition temperature was determined to be 23 K and metamagnetic transition was observed at 2 K\cite{EuHossain}. Upon applying pressure above 3 GPa, Eu exhibit a continuous valence change from a high-temperature divalent state to a low-temperature trivalent state\cite{EuDio}. GdCo$_{2}$Ge$_{2}$ is reported to order at around 40 K\cite{mccall2,Gdnew}. A magnetic x-ray scattering study\cite{GdGood} revealed that this compound orders antiferromagnetically with a temperature-dependent incommensurate wave vector associated with the Gd$^{3+}$ moments that primarily lie in the \textit{ab} plane. Single crystals of TbCo$_{2}$Ge$_{2}$, grown by tri-arc Czochralski method, showed successive metamagnetic transitions at low temperature with commensurate magnetic structure at low applied field and different incommensurate structures at higher fields\cite{Tbshigeoka1,Tbshigeoka}. A neutron diffraction study on TbCo$_{2}$Ge$_{2}$ revealed an incommensurate to commensurate magnetic transition upon cooling\cite{Tbneu,Tbbuschow,TbHo}. In both cases, the magnetic moments are parallel to the \textit{c}-axis. Neutron studies on polycrystalline HoCo$_{2}$Ge$_{2}$\cite{PrHo,TbHo} showed that it orders at 8 K with a ferromagnetic coupling within the plane and an antiferromagnetic coupling between planes. Another study on polycrystalline HoCo$_{2}$Ge$_{2}$ and DyCo$_{2}$Ge$_{2}$ claimed that they both experience an incommensurate to commensurate phase transition with Ho moments lying along \textit{c}-axis while Dy moments deviate from \textit{c}-axis in a temperature-dependent manner\cite{DyHo}. Polycrystalline ErCo$_{2}$Ge$_{2}$ was reported to order antiferromagnetically at 4.2 K with Er moments perpendicular to the \textit{c}-axis\cite{NdEr,Ercp}. Polycrystalline TmCo$_{2}$Ge$_{2}$ was characterized to be ordering at 2.4 K\cite{TmGondek}.

In the present work, a systematic study of magnetic and electric properties of RCo$_{2}$Ge$_{2}$ single crystals, all grown out of high-temperature solutions rich in Ge and Co, is presented for R = Y, La-Nd, Sm-Tm. In addition, specific heat measurements provide further information about transition temperatures as well as qualitative insight into the low temperature degeneracy of the ground state. In this way, all of the studied members of the RCo$_{2}$Ge$_{2}$ series can be systematically compared and contrasted, having been grown and studied under the same conditions. After describing the experimental techniques used in crystal growth and characterization, experimental results will be presented starting with the non-moment-bearing R = Y and La members, combined, and then separately for the each of the other members. After the results section, discussions of trends along the series such as ordering temperature, Curie-Weiss temperature and also on CEF effect will be presented and followed by a brief conclusion.

\section{Experimental}

Single crystals of RCo$_{2}$Ge$_{2}$ were grown using a self flux solution growth method\cite{Canfield92,Canfield01,Canfieldeuro} where CoGe was used as the flux. Typical initial molar ratios of the three elements were R:Co:Ge = 6:47:47 or 8:46:46. The starting elements were added to a 2 ml alumina crucible and sealed in a quartz ampoule under a partial argon atmosphere. The ampoule was then heated up to 1250$^\circ$C and slowly cooled down to around 1100$^\circ$C, at which temperature the excess molten flux was decanted. Given that growth temperatures exceed 1200$^\circ$C, the pressure of the argon in the ampoule was adjusted so as to be as close to atmospheric pressure at 1250$^\circ$C as possible so as to mitigate problems associated with the softening of the silica. All crystals were plate-like with the \textit{c}-axis perpendicular to the plate surface. Despite multiple attempts, crystals of R = Yb and Lu could not be grown.  

\begin{figure}
\begin{center}
\includegraphics[width=75mm, height=61mm]{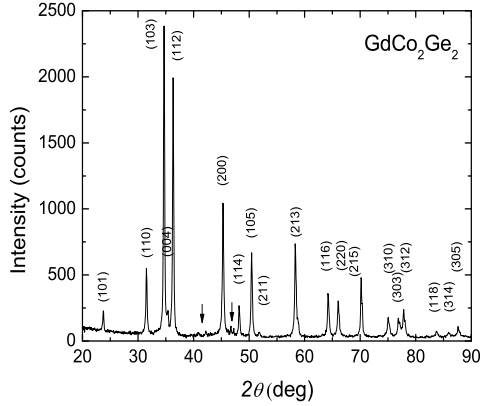}
\end{center}
\caption{Powder x-ray diffraction pattern for GdCo$_{2}$Ge$_{2}$ with (hkl) values for each peak shown. Arrows indicate peaks from the Co-Ge binary phases.}
\label{x-ray}
\end{figure}

The crystal structure was studied at room-temperature using a Rigaku Miniflex powder x-ray diffractometer (Cu K$_{\alpha}$ radiation). Samples were prepared by grinding single crystals into powders, which were then mounted and measured on a single crystal Si, zero-background, sample holder. A typical x-ray pattern is shown in Fig.~\ref{x-ray} for GdCo$_{2}$Ge$_{2}$. All major peaks are consistent with the RCo$_{2}$Ge$_{2}$ tetragonal structure; a minor amount of residual flux gives rise to small peaks associated with Co-Ge binary phases, which are indicated by arrows. Lattice parameters were inferred using GSAS software\cite{GSAS,GUI}. Unit cell parameters are summarized in Table.~\ref{table lattice parameter}, in which the uncertainty is about 0.2\% for refined lattice parameters.

DC magnetization measurements were performed in a Quantum Design Magnetic Property Measurement System (MPMS), superconducting quantum interference device (SQUID) magnetometer (\textit{T} = 1.8 - 350 K, \textit{H}$_{max}$ = 70 kOe). All samples were manually aligned to measure the magnetization along the desired axis. 

Resistivity measurements were performed using a standard 4-probe, AC technique in a Quantum Design Physical Property Measurement System (PPMS) instrument (\textit{f} = 17 Hz, \textit{I} = 1 or 3 mA). For these measurements, samples were cut and polished into rectangular cuboid bars with approximate dimensions of 1.2 $\times$ 0.8 $\times$ 0.3 mm$^{3}$. Epotek-H20E silver epoxy was used to create contacts on all samples and Pt wires were placed such that the current was applied in the \textit{ab} plane. Uncertainty in absolute resistivity due to the measurement of dimension and sample variation is about 20\%.

Heat capacity was measured using the heat capacity option of a Quantum Design PPMS by using the relaxation method in the temperature range of 1.8 - 50 K. In the case of CeCo$_{2}$Ge$_{2}$, a $^{3}$He cooling option was used for measurement down to 0.4 K.

Magnetic transition temperatures are inferred from the maximum of zero-field specific heat, d$\rho$/d\textit{T}\cite{fisherr} and d$\chi_{avg}\textit{T}$/d\textit{T}\cite{fisherxt}, where $\chi_{avg}$ is the polycrystalline averaged magnetic susceptibility calculated from the equation: $\chi_{avg}$ = ($\chi_{c}$ + 2$\chi_{ab}$)/3. Calculated $\chi_{avg}$ is also used in obtaining the effective moment of magnetic rare earth ions in their high temperature paramagnetic state. For CeCo$_{2}$Ge$_{2}$, PrCo$_{2}$Ge$_{2}$, NdCo$_{2}$Ge$_{2}$ and SmCo$_{2}$Ge$_{2}$, the polycrystalline magnetization of LaCo$_{2}$Ge$_{2}$ is subtracted from their polycrystalline magnetization in order to get a more accurate estimate of effective moment, because the temperature-independent magnetization is non-negligible compared with local moment contribution. For the rest of the members in RCo$_{2}$Ge$_{2}$, given that their effective moments increase substantially, the influence of the temperature-independent magnetization is negligible and therefore ignored. Resistivity data was averaged between 3 neighbouring points before taking derivative to achieve better signal-to-noise ratio. The uncertainty in our determinations of the transition temperature was determined by the step width for the specific heat measurement and half width at half maximum for d$\chi\textit{T}$/d\textit{T} and d$\rho$/d\textit{T}. The error bars due to mass uncertainty and different ranges of linear fit is about 2\% for effective moment and 15\% for Curie-Weiss temperature, $\Theta$.   
  
\section{Results}
\subsection{YCo$_{2}$Ge$_{2}$, LaCo$_{2}$Ge$_{2}$}

\begin{figure}[!ht]
\includegraphics[width=130mm, height=100mm]{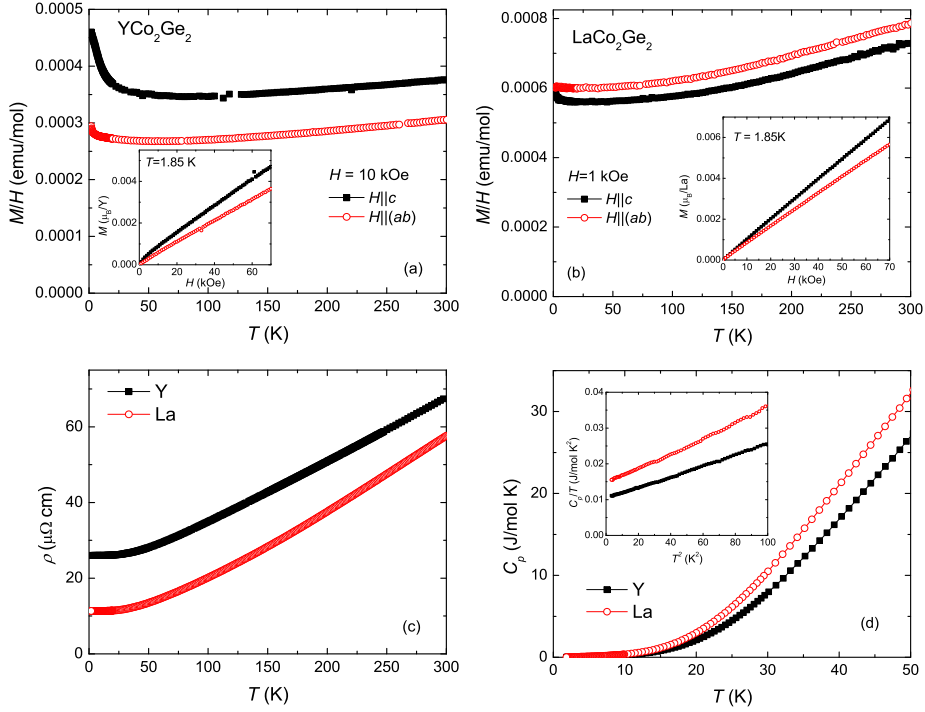}

\caption{(a) Anisotropic magnetic susceptibility of YCo$_{2}$Ge$_{2}$ measured at 10 kOe. (inset: magnetic isotherms measured at 1.85 K) (b) Anisotropic magnetic susceptibility of LaCo$_{2}$Ge$_{2}$ measured at 1 kOe. (inset: magnetic isotherms measured at 1.85 K) (c) Zero-field, in-plane resistivity of YCo$_{2}$Ge$_{2}$ and LaCo$_{2}$Ge$_{2}$. (d) Zero-field specific heat of YCo$_{2}$Ge$_{2}$ and LaCo$_{2}$Ge$_{2}$. (inset: C$_{p}$/\textit{T} versus \textit{T}$^{2}$ at low temperature)}
\label{YR}
\end{figure}

YCo$_{2}$Ge$_{2}$ and LaCo$_{2}$Ge$_{2}$ bear no local moments due to the empty 4f-shells of Y and La. They manifest essentially temperature-independent magnetic susceptibility data, a combination of Pauli-like paramagnetic as well as Landau and core diamagnetic contributions, as shown in Fig.~\ref{YR}. The susceptibility of YCo$_{2}$Ge$_{2}$, above 50 K, stays relatively constant with $\chi_{c} > \chi_{ab}$. The upturns at low temperature can be attributed to very small amount of paramagnetic impurities, most likely one of the rare earth ions with axial anisotropy (e.g. Y$_{1-x}$Tb$_{x}$Co$_{2}$Ge$_{2}$ with x $\sim$ 0.0001 would have a similar sized low-temperature Curie tail). For LaCo$_{2}$Ge$_{2}$, although the upturn at low temperature is smaller than that in YCo$_{2}$Ge$_{2}$, the magnetization does slowly increase with temperature. Compared with its low temperature value, at 300K, the magnetization has increased by nearly 30$\%$. In addition, the magnetic anisotropy for LaCo$_{2}$Ge$_{2}$ is smaller than that of YCo$_{2}$Ge$_{2}$ and of opposite sign with $\chi_{c} < \chi_{ab}$. The magnetic anisotropy in these two compounds, as well as the weak temperature-dependence of the magnetization is probably due to details of their band structure and  Fermi surfaces, which affect the Pauli and Landau contributions to the total magnetization. Magnetic isotherms measured on YCo$_{2}$Ge$_{2}$ and LaCo$_{2}$Ge$_{2}$ at 1.85 K are approximately linear for both \textit{H}$\parallel\textit{c}$ and \textit{H}$\parallel\textit{(ab)}$. Slight curvature is consistent with small amounts of paramagnetic impurities.

The zero-field resistivity of YCo$_{2}$Ge$_{2}$ and LaCo$_{2}$Ge$_{2}$ show typical, metallic behavior, without any anomaly observed down to 1.8 K. The residual resistance ratios (RRR $\equiv \rho(300 K)/\rho(2 K)$) for YCo$_{2}$Ge$_{2}$ and LaCo$_{2}$Ge$_{2}$ are 2.7 and 5.0 respectively.

The temperature-dependent specific heat C$_{p}$ data are similar for these two compounds. The estimated Debye temperatures are about 400 K for YCo$_{2}$Ge$_{2}$ and 370 K for LaCo$_{2}$Ge$_{2}$. The relative value of two Debye temperatures roughly follows what would be predicted by Debye model with the different molecular masses associated with the change from Y to La. The linear coefficients of specific heat, $\gamma$, extracted from the plot of C$_{p}$/\textit{T} versus \textit{T}$^{2}$, are 10.4 mJ/mol K$^{2}$ for YCo$_{2}$Ge$_{2}$ and 14.6 mJ/mol K$^{2}$ for LaCo$_{2}$Ge$_{2}$. In calculating the magnetic entropies of RCo$_{2}$Ge$_{2}$ magnetic members, although YCo$_{2}$Ge$_{2}$ and LaCo$_{2}$Ge$_{2}$ can provide an estimation of the non-magnetic specific heat, at a quantitative level, for this series, they do not allow for a detailed calculation, as the magnetic entropy of GdCo$_{2}$Ge$_{2}$ does not reach expected \textit{R}ln8 upon ordering, indicating that this approximation is not ideal for this system. At a qualitative level, however, it is still interesting to get a sense of how CEF splitting reduces the free ion's degeneracy for the members that manifest clear magnetic anisotropy. The specific heat of LaCo$_{2}$Ge$_{2}$ was used to approximate the non-magnetic contribution across the series.

\subsection{CeCo$_{2}$Ge$_{2}$}
\begin{figure}[!ht]
\includegraphics[width=130mm, height=100mm]{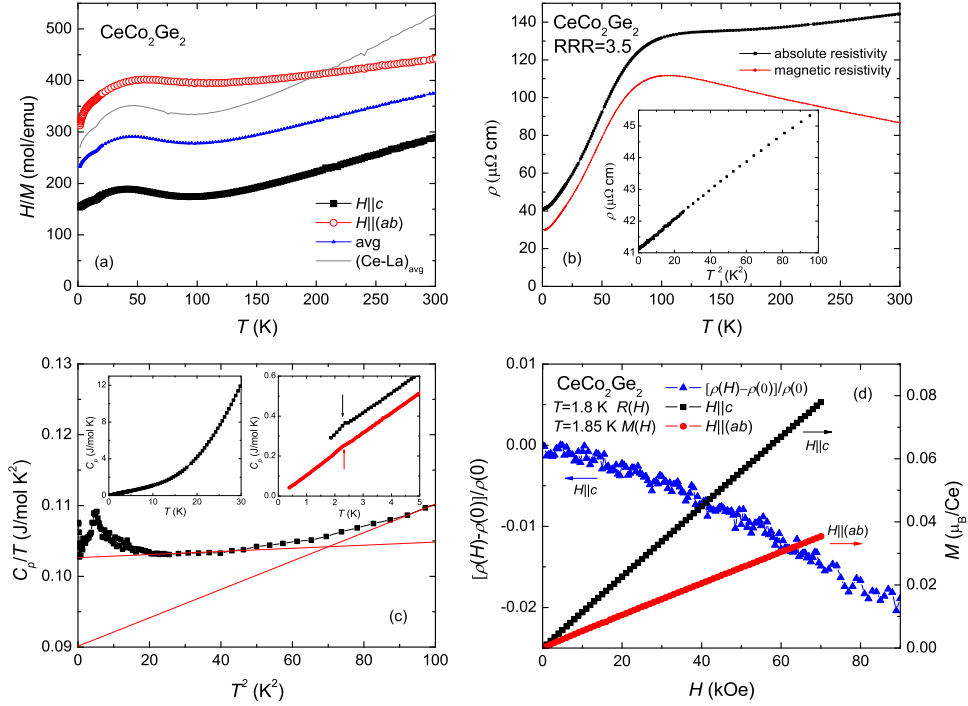}

\caption{Measurements of CeCo$_{2}$Ge$_{2}$ (a) Anisotropic and polycrystalline averaged inverse magnetic susceptibility measured at 1 kOe. The method used in calculating polycrystalline averaged magnetic susceptibility was described in the "Experimental" section. Grey line is the polycrystalline averaged data corrected by LaCo$_{2}$Ge$_{2}$ (see text). (b) Zero-field resistivity (Black) and magnetic component of resistivity ($\rho_{mag} = \rho_{CeCo_2Ge_2} - \rho_{LaCo_2Ge_2}$ shown in Red) (inset: low-temperature resistivity versus $\textit{T}^{2}$) (c) Zero-field C$_{p}/\textit{T}$ versus \textit{T}$^{2}$. Red lines indicate the range of $\gamma$ in the temperature range of 5-10 K. (left inset: zero-field specific heat. right inset: low temperature specific heat from two different samples with the anomaly indicated by arrows. The black data is shifted upward by 0.1 J/mol K$^{2}$ to avoid overlapping) (d) Magnetization isotherms and normalized magnetoresistance.}
\label{Ce}
\end{figure}

Fig.~\ref{Ce} summarizes the thermodynamic and transport measurement results for CeCo$_{2}$Ge$_{2}$. Temperature-dependent magnetization of CeCo$_{2}$Ge$_{2}$ measured at 1 kOe is anisotropic with $\chi_{c} > \chi_{ab}$. The inverse magnetic susceptibility shows a broad minimum at around 100 K, a maximum near 40 K and a lower temperature drop that is likely associated with an impurity tail, which is similar to earlier work\cite{Cefujii}. Although the weak variation of magnetic susceptibility was explained by the Ce ion being in an intermediate valence state\cite{Cefujii}, given that the overall signal of magnetic susceptibility is small, evaluating the effective moment without taking the non-local-moment contribution into account may not be valid. The polycrystalline averaged susceptibility of LaCo$_{2}$Ge$_{2}$ was subtracted from that of CeCo$_{2}$Ge$_{2}$ in order to estimate the effective moment of Ce, as shown in the gray curve in Fig. ~\ref{Ce}(a). A $\mu_{eff}$ value of 2.6 $\mu_{B}$ is inferred above 200 K, which is close to the theoretical value of Ce$^{3+}$ (2.5 $\mu_{B}$). Therefore, at high temperatures Ce is probably close to being in its trivalent state in CeCo$_{2}$Ge$_{2}$. This interpretation is consistent with a near-edge x-ray absorption study\cite{Cex}. In \textit{H}/\textit{M}(\textit{T}), there seems to exist a small kink at around 15 K. However, as it is not a typical feature for magnetic ordering and no clear feature was observed in other measurements, it is not clear at this stage what physical meaning it has. 

The zero-field resistivity of CeCo$_{2}$Ge$_{2}$ shows little change for 100 - 300 K. A broad shoulder is observed at 100 K and is followed by a much stronger drop at lower temperatures. The RRR for CeCo$_{2}$Ge$_{2}$ is 3.6. At low temperature, its resistivity is proportional to \textit{T}$^{2}$ as shown in the inset of Fig.~\ref{Ce}(b) with a slope A = 0.045 $\mu\Omega$ cm/K$^2$. Magnetic resistivity ($\rho_{m}$ = $\rho_{Ce}$ - $\rho_{La}$) shows a broad maximum centered around 100 K. 

The zero-field specific heat data measured down to 0.4 K is shown in Fig.~\ref{Ce}(c). In the low temperature region, there is an anomaly in the C$_{p}/\textit{T}$ versus $\textit{T}^{2}$ plot at around 2.3 K, which was not observed in previous $^{3}$He measurements\cite{takashi}. However, no similar feature was observed in either resistivity or in magnetic susceptibility data. It may be because, at such a low temperature, neither resistivity nor magnetization measurements had high enough resolution to resolve the signatures associated with the feature seen in the specific heat. This feature may indicate some interesting physics that we do not yet understand and ultimately may merit further investigation. It is worth noting here that in the $^{4}He$ C$_{p}$ versus \textit{T} data, the feature was clearly seen albeit at the end of the measurement range (Fig.~\ref{Ce}(c)).

The electronic specific heat, $\gamma$, of CeCo$_{2}$Ge$_{2}$ is $\sim$ 90-103 mJ/mol K$^{2}$ in the range of 5-10 K, which is much larger than that of YCo$_{2}$Ge$_{2}$ and LaCo$_{2}$Ge$_{2}$ (see two estimates as red lines in Fig.~\ref{Ce}(c)). This is consistent with a low temperature, Kondo screened state. In the perspective of entropy removal, the Kondo temperature \textit{T}$_{K}$ can be roughly estimated by \textit{T}$_{K}$ = \textit{R} ln(N)/$\gamma$, where N is the degeneracy of the CEF split ground state being Kondo screened, and \textit{R} is the ideal gas constant. \textit{T}$_{K}$ estimated in this way is about 60 K if N = 2 is used and 120 K if N = 4 is used, which, qualitatively, are consistent with the maximum observed in magnetic resistivity. The choice of degeneracy agrees with generalized Kadowaki-Woods relation\cite{kw}, in which CeCo$_{2}$Ge$_{2}$ lies in between N = 2 and N = 4. Numerical solution of Coqblin-Schrieffer model\cite{Rajan} can also be applied to fit magnetic susceptibility data of CeCo$_{2}$Ge$_{2}$, yet it gives a much higher Kondo temperature (\textit{T}$_{K} \sim$ 240 K). The discrepancy in \textit{T}$_{K}$ may come from a poor non-magnetic background subtraction.

The field-dependent magnetization measured at 1.85 K is anisotropic and increases linearly with field up to 70 kOe. Magnetoresistance measured at 1.8 K with \textit{H}$\parallel$\textit{c} is weak, negative and appears to be non-linear in applied field. 

\subsection{PrCo$_{2}$Ge$_{2}$}
\begin{figure}[!ht]
\includegraphics[width=130mm, height=100mm]{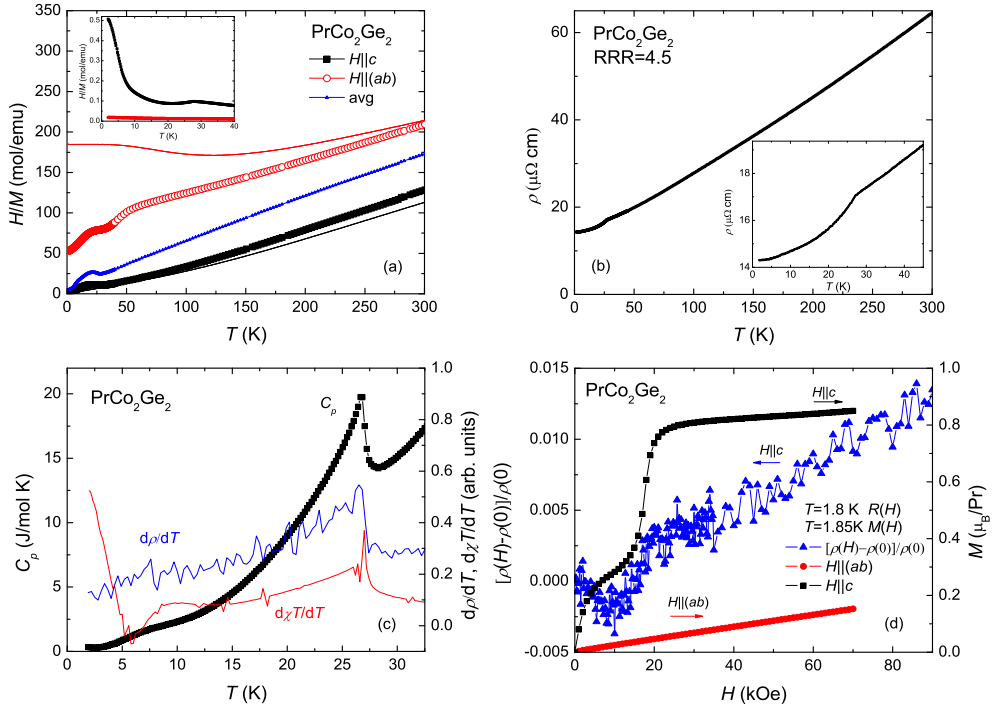}

\caption{Measurements of PrCo$_{2}$Ge$_{2}$ (a) Anisotropic and polycrystalline averaged inverse magnetic susceptibility measured at 1 kOe. Solid curves represent magnetic susceptibilities calculated from model based on CEF parameter. (See Discussion section) (inset: low-temperature magnetic susceptibility) (b) Zero-field resistivity (inset: low-temperature resistivity) (c) Zero-field specific heat on the left scale, d$\chi$\textit{T}/d\textit{T} and d$\rho$/d\textit{T} on the right with arbitrary units. (d) Magnetization isotherms and normalized magnetoresistance.}
\label{Pr}
\end{figure}

The magnetic susceptibility of PrCo$_{2}$Ge$_{2}$ in Fig.~\ref{Pr}(a) is also anisotropic with $\chi_{c} > \chi_{ab}$. A single transition is seen in the \textit{M}/\textit{H}(\textit{T}) data at around 27 K. Since the magnetization is still increasing with decreasing temperature at the lowest temperature we could reach, the ordered state of PrCo$_{2}$Ge$_{2}$ may have a net ferromagnetic component and/or PrCo$_{2}$Ge$_{2}$ may have further, T $<$ 1.8 K, transitions as well. At high temperatures, Curie-Weiss behavior is clear with $\Theta_{(ab)}$ = -180 K, $\Theta_{c}$ = -38 K and $\Theta_{ave}$ = -23 K. The effective moment obtained from its polycrystalline average in the paramagnetic state is 3.5 $\mu_{B}$ if the temperature-independent contribution represented by LaCo$_{2}$Ge$_{2}$ is subtracted, which is close to the theoretical prediction for free Pr$^{3+}$ ion (3.6 $\mu_{B}$). (Note: These values are compiled for all measured materials in Table~\ref{all data} below)

The temperature-dependent, zero-field resistivity data (Fig.~\ref{Pr}(b)) show a clear loss of spin-disorder scattering feature associated with the transition temperature obtained from magnetic susceptibility measurements. The RRR is 4.5. 

In Fig.~\ref{Pr}(c), the plots of d$\chi\textit{T}$/d\textit{T} as well as d$\rho$/d\textit{T} are shown in arbitrary units together with zero-field specific heat data. Features associated with an antiferromagnetic phase transition are consistent in all three measurements with the transition temperature inferred as 26.7 $\pm$ 0.9 K. The specific heat data also reveal that there exists a broad shoulder around 8 K, which had been reported in an earlier study\cite{Prvej}. This C$_{p}$(T) shoulder occurs at the same temperature range as the low-temperature upturn in the magnetic susceptibility and is probably associated with changes in the magnetic excitation spectrum. Estimated magnetic entropy indicates that the ordering moment comes from a pseudo-doublet ground state. 

In Fig.~\ref{Pr}(d), the magnetization isotherm measured at \textit{T} = 1.85 K increases linearly when \textit{H}$\parallel$\textit{(ab)}. For \textit{H}$\parallel$\textit{c}, metamagnetic transitions were observed with magnetization showing step-like behavior with two well defined plateaus at $\sim$ 0.2 $\mu_{B}$ and $\sim$ 0.8 $\mu_{B}$. This is consistent with observed metamagnetic transition at 20 kOe\cite{Prvej}. Although we didn't observe a clear magnetic hysteresis at 1.85 K, the first plateau at 0.2 $\mu_{B}$ probably represents a small ferromagnetic component in the intrinsic ordered state given the sharp increase of magnetization before the plateau and the linearity of magnetoresistance within the field range. Another metamagnetic transition at around 100 kOe was also reported\cite{Prvej}, above which Pr$^{3+}$ is fully saturated. Magnetoresistance measured at 1.8 K with \textit{H}$\parallel$\textit{c} shows consistent changes at the same critical fields, which was not well resolved in the above mentioned work.\cite{Prvej}

\subsection{NdCo$_{2}$Ge$_{2}$}

\begin{figure}[!ht]
\includegraphics[width=130mm, height=100mm]{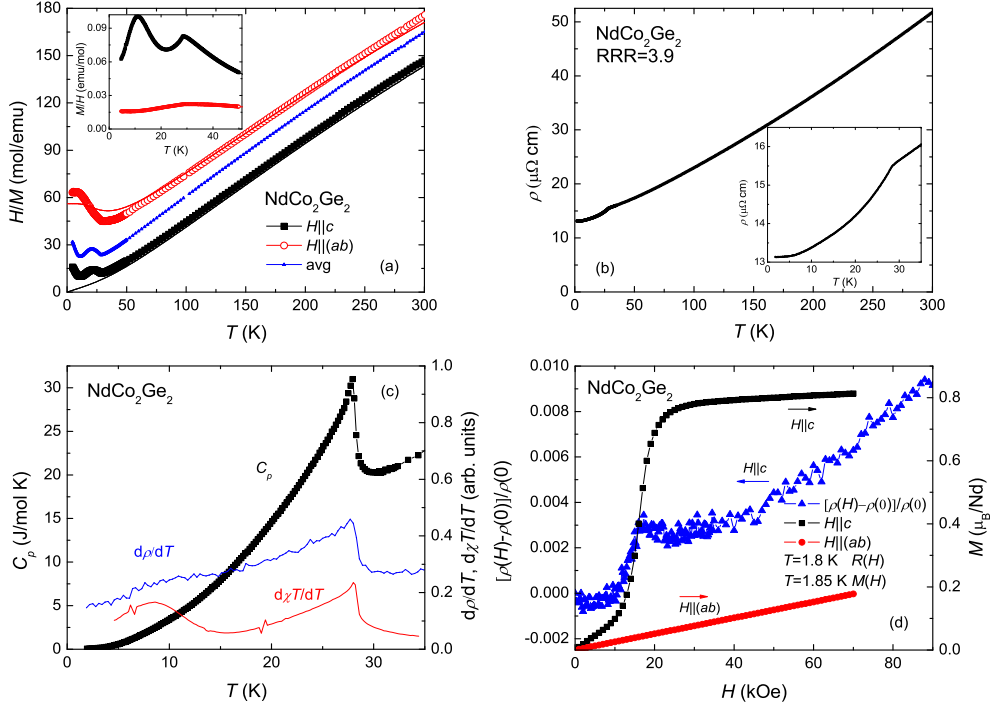}

\caption{Measurements of NdCo$_{2}$Ge$_{2}$ (a) Anisotropic and polycrystalline averaged inverse magnetic susceptibility measured at 1 kOe. Solid curves represent magnetic susceptibilities calculated from model based on CEF parameter. (inset: low-temperature magnetic susceptibility) (b) Zero-field resistivity (inset: low-temperature resistivity) (c) Zero-field specific heat on the left scale, d$\chi$\textit{T}/d\textit{T} and d$\rho$/d\textit{T} in arbitrary units. (d) Magnetization isotherms and normalized magnetoresistance.}
\label{Nd}
\end{figure}

The temperature-dependent magnetization of NdCo$_{2}$Ge$_{2}$ (Fig.~\ref{Nd}(a)) is anisotropic with $\chi_{c} > \chi_{ab}$. There are two low temperature features in the \textit{M}/\textit{H}(\textit{T}) data: a sharp feature, with a transition temperature at around 28 K and a broad feature near 9 K. Under higher applied fields, the two maxima were reported to gradually shift to lower temperatures with the second magnetic transition being almost smoothed out above 50 kOe\cite{Cefujii}. A high temperature, Curie-Weiss fit to the polycrystalline averaged data suggests the effective moment to be 3.6 $\mu_{B}$, which is the same as expected for Nd$^{3+}$ (3.6 $\mu_{B}$). The paramagnetic Curie-Weiss temperatures are $\Theta_{(ab)}$ = -50 K, $\Theta_{c}$ = 11 K and $\Theta_{ave}$ = -19 K. 

Fig.~\ref{Nd}(b) shows the zero-field resistivity data of NdCo$_{2}$Ge$_{2}$ with RRR = 3.9. Only one transition was observed, which is close to the ordering temperature obtained above. A rather broad and smeared shoulder may correspond to a second magnetic transition. However, the feature in d$\rho$/d\textit{T} is too broad to offer any quantitative information. 

The specific heat measurement also shows only one signature of phase transition, which confirms the above obtained  ordering temperature. This is in agreement with reported specific heat measurement\cite{Ndcp}. The transition temperature associated with the clear high-temperature feature is 27.9 $\pm$ 0.9 K. Estimated magnetic entropy indicates that there is \textit{R}ln4 entropy removed below the magnetic ordering temperature.

There have been discussions on the nature of the feature observed in temperature-dependent magnetization data at around 9 K\cite{Cefujii,NdEr,Ndcp,Ndandre}. It was speculated that the feature only emerges under certain finite applied fields. However, at 10 Oe, both the temperature and the relative size of two maxima in \textit{M}/\textit{H}(\textit{T}) remain unchanged (not shown here). This feature perhaps originates from a change in magnon spectrum due to thermal depopulation of split CEF levels. If there really is a magnetic transition, then, in the perspective of specific heat measurement, the absence of a clear signature indicates that it is a transition that does not involve any significant amount of magnetic entropy.

The magnetization isotherms measured at 1.85 K (Fig.~\ref{Nd}(d)) are anisotropic. When \textit{H}$\parallel$\textit{(ab)}, the magnetization increases linearly with applied field whereas when \textit{H}$\parallel$\textit{c}, a metamagnetic transition at around 17 kOe was observed and followed by a plateau at above 30 kOe. This is consistent with an early polycrystalline study\cite{NdEr}. But in this case, the transition is much sharper with a better defined critical field. Another metamagnetic transition at about 110 kOe was reported for polycrystalline sample, above which the magnetization is approaching that of Nd$^{3+}$'s saturated value\cite{Prvin}. Magnetoresistance measured at 1.8 K with magnetic field parallel to \textit{c}-axis shows a local maximum around the same critical field where metamagnetic transition was observed. After this, the resistivity increases linearly with applied field above 50 kOe.

\subsection{SmCo$_{2}$Ge$_{2}$}

\begin{figure}[!ht]
\includegraphics[width=130mm, height=100mm]{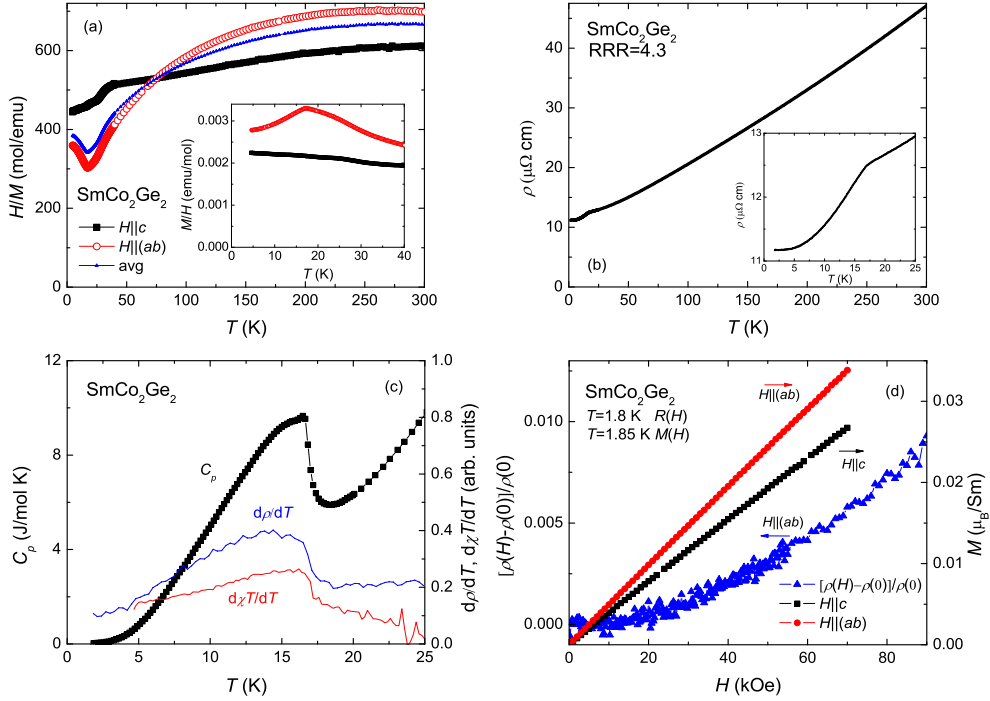}

\caption{Measurements of SmCo$_{2}$Ge$_{2}$ (a) Anisotropic and polycrystalline averaged inverse magnetic susceptibility measured at 1 kOe. (inset: low-temperature magnetic susceptibility) (b) Zero-field resistivity (inset: low-temperature resistivity) (c) Zero-field specific heat on the left scale, d$\chi$\textit{T}/d\textit{T} and d$\rho$/d\textit{T} in arbitrary units. (d) Magnetization isotherms and normalized magnetoresistance.}
\label{Sm}
\end{figure}

As shown in Fig.~\ref{Sm}(a), the temperature-dependent magnetization of SmCo$_{2}$Ge$_{2}$ is quite different from previous members. It is anisotropic but its anisotropy changes at around 75 K, below which $\chi_{c} < \chi_{ab}$. In addition, its inverse magnetic susceptibility does not follow Curie-Weiss behavior up to 300 K, but instead, shows a wide negative curvature and tends to saturate at room temperature. This behavior is commonly seen in Sm based compounds\cite{RNi, RSb2, RAgSb2} and could come from thermal-population of Sm$^{3+}$'s Hund's rule excited states or a fluctuation of Sm's valent state. A clear feature of an antiferromagnetic transition can be seen at around 17 K.

The temperature-dependent resistivity is showed in Fig.~\ref{Sm}(b) with RRR = 4.3. A well pronounced signature of the loss of spin disorder scattering can be seen at the transition temperature. 

The zero-field, temperature-dependent specific heat does not exhibit a typical $\lambda$ shape, with a broad curvature sitting approximately 1 K below the transition temperature inferred at 16.5 $\pm$ 0.2 K. The consistency of peak shape in C$_{p}$(\textit{T}), d$\chi\textit{T}$/d\textit{T} and d$\rho$/d\textit{T} near the transition temperature may also be an indication of a cascade of closely spaced transitions that can not be well resolved by the present measurements. It is also possible that a specific type of incommensurate magnetic ordering structure leads to this broadened feature\cite{Blanco}. Magnetic entropy suggests that the ordered state evolves out of a CEF split, doublet state. 

The magnetization isotherms (Fig. ~\ref{Sm}(d)) are linear with applied field up to 70 kOe with \textit{H}$\parallel$\textit{(ab)} yielding a slightly larger slope. The magnetoresistance also increases as applied field increases in a quadratic manner. No clear feature associated with a metamagnetic transition was observed.

\subsection{EuCo$_{2}$Ge$_{2}$}

\begin{figure}[!ht]
\includegraphics[width=130mm, height=100mm]{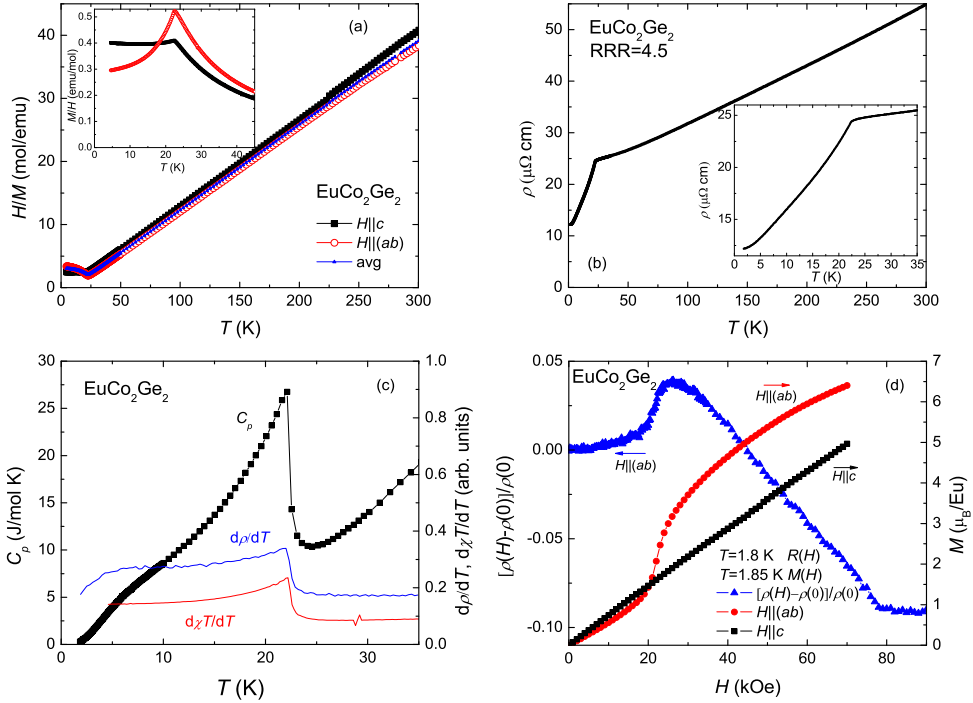}

\caption{Measurements of EuCo$_{2}$Ge$_{2}$ (a) Anisotropic and polycrystalline averaged inverse magnetic susceptibility measured at 1 kOe. (inset: low-temperature magnetic susceptibility) (b) Zero-field resistivity (inset: low-temperature resistivity) (c) Zero-field specific heat on the left scale, d$\chi$\textit{T}/d\textit{T} and d$\rho$/d\textit{T} in arbitrary units. (d) Magnetization isotherms and normalized magnetoresistance.}
\label{Eu}
\end{figure}

Fig.~\ref{Eu} presents the measured data on EuCo$_{2}$Ge$_{2}$. The temperature-dependent magnetization of EuCo$_{2}$Ge$_{2}$ is essentially isotropic above 50 K. In its high temperature paramagnetic state, a linear fit on inverse magnetic susceptibility gives $\Theta_{ave}$ = 10 K and $\mu_{eff}$ = 7.7 $\mu_{B}$, which suggests that Eu is in a divalent state. This is consistent with the anomalously large unit cell volume of EuCo$_{2}$Ge$_{2}$ shown in Table.~\ref{table lattice parameter} and Fig.~\ref{lattice} below. On the one hand, a positive Curie-Weiss temperature seems to indicate a ferromagnetic exchange interaction between Eu$^{2+}$ ions. On the other hand, a single transition temperature at 22.2 $\pm$ 0.3 K was inferred from d$\chi\textit{T}/d\textit{T}$, and the transition appears to be antiferromagnetic with the ordered moments being  perpendicular to the crystallographic c-axis, given essentially temperature-independent $M/H(T)$ below T$_{N}$ for $\textit{H} \parallel \textit{c}$.

In Fig.~\ref{Eu}(b), the zero-field resistivity with RRR = 4.5 also indicates a single transition at around 22 K. A clear decrease in the resistivity can be seen at the transition temperature. The zero-field temperature-dependent specific heat data, shown in Fig.~\ref{Eu}(c), yield a consistent transition temperature. Combining three different measurements, the magnetic transition temperature of EuCo$_{2}$Ge$_{2}$ is  22.1 $\pm$ 0.4 K. In a recent study on Eu-based intermetallic compound, EuNiGe$_{3}$\cite{Johnston}, a similar concomitance of a positive Curie-Weiss temperature and antiferromagnetic transition was observed and explained by a collinear A-type antiferromagnetic structure. Similarly, in the current case, given the anisotropy below T$_{N}$, it is possible that there exist a ferromagnetic interaction along \textit{c} and an antiferromagnetic interaction within the \textit{ab} plane, namely C-type antiferromagnetic structure. It will be interesting to study the magnetic structure of EuCo$_{2}$Ge$_{2}$ via magnetic x-ray scattering, since both Eu and Gd have similar ordered moment directions and yet different signs of $\Theta_{avg}$ (Data for GdCo$_{2}$Ge$_{2}$ will be shown in the next section). As zero CEF splitting for Eu$^{2+}$ Hund's rule ground state multiplets would predict, the magnetic entropy calculated for this compound reaches \textit{R}ln8 at the ordering temperature.

The magnetization isotherms of EuCo$_{2}$Ge$_{2}$ at \textit{T} = 1.85 K are anisotropic. When \textit{H}$\parallel$\textit{c}, the magnetization increases linearly with applied field. For \textit{H}$\parallel$\textit{(ab)}, below 20 kOe, the magnetization increases linearly with applied field with a slightly smaller slope compare with that in \textit{H}$\parallel$\textit{c}. At around 20 kOe, the magnetization in \textit{ab} plane suddenly increases, after which it shows a weak negative curvature, similar to the earlier reported features\cite{EuHossain}. Magnetoresistance also increases at around 20 kOe and manifests a local maximum at around 25 kOe. It then decreases as applied field increases. At 80 kOe, the magnetoresistance manifests another sharp change in slope, which may indicate another metamagnetic transition. Further study on magnetization at higher field may help to clarify this point.

\subsection{GdCo$_{2}$Ge$_{2}$}

\begin{figure}[!ht]
\includegraphics[width=130mm, height=100mm]{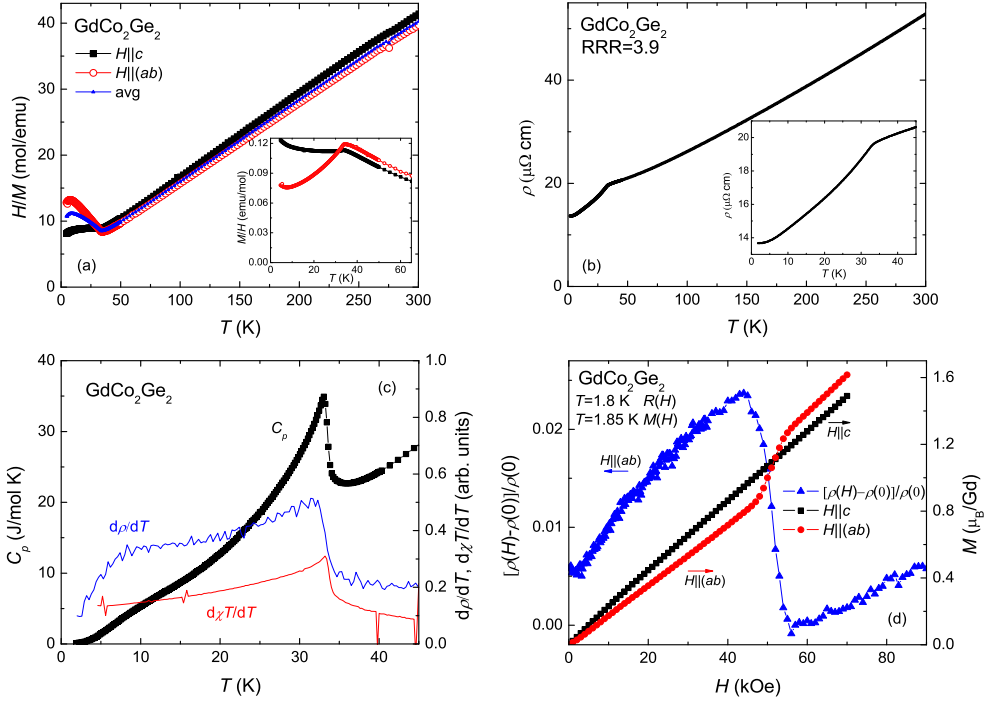}

\caption{Measurements of GdCo$_{2}$Ge$_{2}$ (a) Anisotropic and polycrystalline averaged inverse magnetic susceptibility measured at 1 kOe. (inset: low-temperature magnetic susceptibility) (b) Zero-field resistivity (inset: low-temperature resistivity) (c) Zero-field specific heat on the left scale, d$\chi$\textit{T}/d\textit{T} and d$\rho$/d\textit{T} in arbitrary units. (d) Magnetization isotherms and normalized magnetoresistance.}
\label{Gd}
\end{figure}

The temperature-dependent magnetization of GdCo$_{2}$Ge$_{2}$, shown in Fig.~\ref{Gd}(a), resembles that of EuCo$_{2}$Ge$_{2}$. It is also isotropic as expected for a Hund's rule ground state multiplet J with zero angular momentum. A high temperature Curie-Weiss fit on polycrystalline averaged \textit{H}/\textit{M} gives a negative $\Theta_{ave}$ = -32 K, which is consistent with simple antiferromagnetic ordering. The effective moment inferred is 8.1 $\mu_{B}$, close to the the theoretical value of 7.9 $\mu_{B}$/Gd$^{3+}$. Only one transition at about 33 K was observed. The anisotropy on $M/H(T)$ below T$_{N}$, as well as the magnetic x-ray scattering results\cite{GdGood} indicate that the ordered moments are in the basal, \textit{ab} plane. 

In Fig.~\ref{Gd}(b), the temperature-dependent resistivity also shows a clear loss of spin-disorder scattering at the transition temperature. The RRR is 4.9. 

The zero-field specific heat of GdCo$_{2}$Ge$_{2}$ (Fig. ~\ref{Gd}(c)) shows a clear signature of transition. Summarizing the data from $\rho$(\textit{T}), \textit{M}/\textit{H}(\textit{T}) and C$_{p}$(\textit{T}), the magnetic transition temperature of GdCo$_{2}$Ge$_{2}$ is 31.7 $\pm$ 1.5 K, 33.2 $\pm$ 1.8 K, 33.1 $\pm$ 0.2 K respectively resulting in an average value of 32.6 $\pm$ 2.4 K, which is smaller compared to earlier reports on polycrystalline materials: 37.5 K\cite{Gdnew} and 40 K\cite{mccall2}, but fully consistent with the single crystal work reported\cite{GdGood}. In addition, as can be seen in the magnetic specific heat of GdCo$_{2}$Ge$_{2}$ as well as in EuCo$_{2}$Ge$_{2}$, there is a hump showing up at around 25$\%$ of their ordering temperatures. In studies of specific heat of Gd based compounds\cite{Blanco,Bouvier}, this phenomenon was observed to be common and was explained to arise naturally from mean field calculation for a (2J+1)-fold degenerate multiplets. Although the study was specifically focusing on Gd, it is not surprising that the specific heat of EuCo$_{2}$Ge$_{2}$ can be explained in the same way, as Eu is in its divalent state. 

At 1.85 K, the field-dependent magnetization is similar to that of EuCo$_{2}$Ge$_{2}$. When \textit{H}$\parallel$\textit{c}, the magnetization is roughly linear in \textit{H} whereas when \textit{H}$\parallel$\textit{(ab)}, a metamagnetic transition occurs around 52 kOe.  Although \textit{ab} plane is not a well defined direction, 52 kOe, the field value found for GdCo$_{2}$Ge$_{2}$, is more than double the value of 20 kOe found for EuCo$_{2}$Ge$_{2}$. Although simple geometric arguments could give a factor of $\sqrt{2}$ in critical field associated with in-plane orientation, so a factor of $\sim$2.5 makes the transition field in GdCo$_{2}$Ge$_{2}$ unambiguously higher than that in EuCo$_{2}$Ge$_{2}$. Magnetoresistence measured at 1.8 K suddenly decreases at the critical field of the metamagnetic transition and increases both before and after the transition with different rates.

\subsection{TbCo$_{2}$Ge$_{2}$}

\begin{figure}[!ht]
\includegraphics[width=130mm, height=100mm]{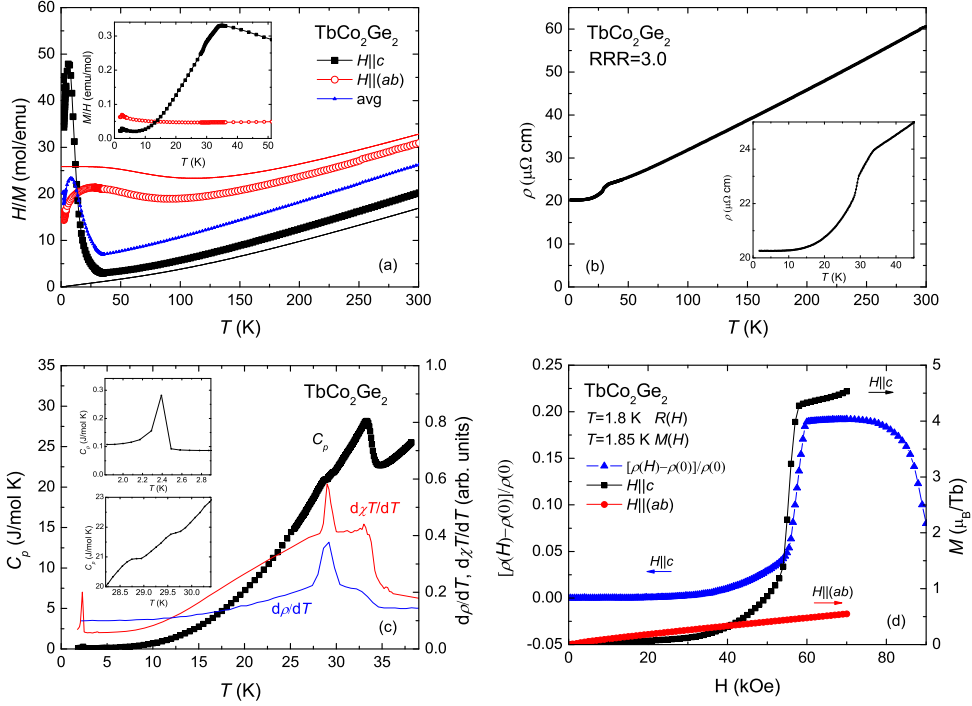}

\caption{Measurements of TbCo$_{2}$Ge$_{2}$ (a) Anisotropic and polycrystalline averaged inverse magnetic susceptibility measured at 1 kOe. Solid curves represent magnetic susceptibilities calculated from model based on CEF parameter. (inset: low-temperature magnetic susceptibility) (b) Zero-field resistivity (inset: low-temperature resistivity) (c) Zero-field specific heat on the left scale, d$\chi$\textit{T}/d\textit{T} and d$\rho$/d\textit{T} in arbitrary units. (insets: zoom-in view of C$_{p}$(\textit{T}) at 2.4 K and 29.2 K) (d) Magnetization isotherms and normalized magnetoresistance.}
\label{Tb}
\end{figure}

The temperature-dependent magnetization of TbCo$_{2}$Ge$_{2}$ is strongly anisotropic with $\chi_{c}>\chi_{ab}$ (Fig.~\ref{Tb}(a)). The sign of Curie-Weiss temperatures varies with the orientation: $\Theta_{(ab)}$ = -160 K, $\Theta_{c}$ = 37 K and $\Theta_{ave}$ = -33 K. The value of effective moment is 9.9 $\mu_{B}$, which is close to the theoretical value of Tb$^{3+}$ (9.7 $\mu_{B}$). Multiple transitions were observed, including two successive transitions at around 33 K and 29 K and a low temperature transition at 2.3 K. The ordering temperature is comparable with that of GdCo$_{2}$Ge$_{2}$, which contradicts the prediction of simple de Gennes scaling. This will be expanded upon later, in the discussion section. 

A clear loss of spin-disorder scattering can be seen in the zero-field resistivity measurement as shown in Fig.~\ref{Tb}(b). RRR is 3.0 for this compound. The two higher temperature transitions seen in the magnetic susceptibility measurement are also clearly visible in the resistivity data, whereas no anomaly was observed at around 2.3 K.

The zero-field specific heat measurement is shown in Fig.~\ref{Tb}(c) and its insets. There are four transitions at 33.3 K, 29.7 K, 28.8 K and 2.4 K. Two transitions show up in the temperature range around the second transition observed in \textit{M}/\textit{H}(\textit{T}) data. It is possible that our magnetic susceptibility measurement simply does not resolve those two closely spaced transitions. An alternative scenario could be that those two transitions actually merge under a small finite applied field. To test this assumption, a temperature-dependent magnetization was measured at 200 Oe (not shown here). Even at this low field, no splitting of the transition is seen around the temperature range of interest. In Table.~\ref{all data}, the transition temperature inferred from C$_{p}$(\textit{T}) will be used for the two middle neighbouring transitions. The magnetic phase diagram constructed for TbCo$_{2}$Ge$_{2}$ by Shigeoka \textit{et al}\cite{Tbshigeoka1,Tbshigeoka} indicates that, in zero applied field, there is only one magnetic phase below T$_{N}$. However, in the present study, up to 4 different phases are found above 1.8 K without applied field. The estimated magnetic entropy reached \textit{R}ln2 at the ordering temperature.

The magnetization isotherms measured at 1.85 K show a metamagnetic transition with a step-like behavior when \textit{H}$\parallel$\textit{c}. A plateau at about 4.5 $\mu_{B}$/Tb$^{3+}$ was reached after the transition. In-plane field-dependent magnetization increases linearly with applied field. Magnetoresistance shows one more metamagnetic transition at around 80 kOe. Both metamagnetic transition fields are consistent with previously reported work\cite{Tbshigeoka1,Tbshigeoka,TbDy}. Another reported metamagnetic transition at 69 kOe\cite{Tbshigeoka1,Tbshigeoka} was not clearly observed in our measurement, although the broad decrease in magnetoresistance above 70 kOe may actually reflect two metamagnetic transitions as reported. 

\subsection{DyCo$_{2}$Ge$_{2}$}

\begin{figure}[!ht]
\includegraphics[width=130mm, height=100mm]{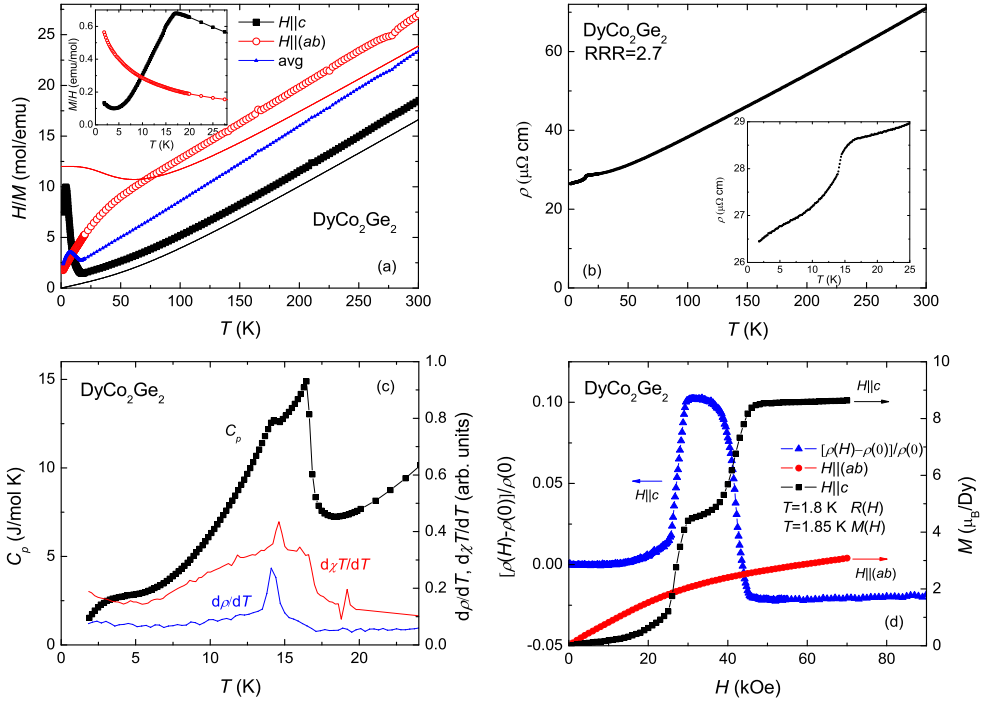}

\caption{Measurements of DyCo$_{2}$Ge$_{2}$ (a) Anisotropic and polycrystalline averaged inverse magnetic susceptibility measured at 1 kOe. Solid curves represent magnetic susceptibilities calculated from model based on CEF parameter. (inset: low-temperature magnetic susceptibility) (b) Zero-field resistivity (inset: low-temperature resistivity) (c) Zero-field specific heat on the left scale, d$\chi$\textit{T}/d\textit{T} and d$\rho$/d\textit{T} in arbitrary units. (d) Magnetization isotherms and normalized magnetoresistance.}
\label{Dy}
\end{figure}

The temperature-dependent magnetization of DyCo$_{2}$Ge$_{2}$ (Fig.~\ref{Dy}(a)) is similar to that of TbCo$_{2}$Ge$_{2}$. Two transitions at about 16 K and 14 K were observed. Curie-Weiss temperatures inferred in its paramagnetic state are $\Theta_{(ab)}$ = -80 K, $\Theta_{c}$ = 28 K and $\Theta_{ave}$ = -16 K. The effective moment is 10.4 $\mu_{B}$, which is close to the theoretical value for Dy$^{3+}$ (10.6 $\mu_{B}$). 

The temperature-dependent, zero-field resistivity of DyCo$_{2}$Ge$_{2}$ has a RRR of 2.7. Two clear features can be seen in the inset of Fig.~\ref{Dy}(b). Around the transition temperatures obtained above, there are very clear changes in slope.

The temperature-dependent specific heat of DyCo$_{2}$Ge$_{2}$ also shows two transitions. Together with the values obtained above, magnetic transition temperatures for DyCo$_{2}$Ge$_{2}$ are 16.0 $\pm$ 0.7 K and 14.4 $\pm$ 0.4 K (Fig.~\ref{Dy}(c)). The estimated magnetic entropy at the ordering temperature is close to \textit{R}ln4. 

Clear features were observed in the magnetic isotherms measured at 1.85 K as shown in Fig.~\ref{Dy}(d). For \textit{H}$\parallel$\textit{(ab)}, a negative curvature was observed. For \textit{H}$\parallel$\textit{c}, which is the easy axis, the magnetization shows sudden jumps between different, well-defined, plateaus at around 25 kOe and 40 kOe. Although it tends to saturate after about 50 kOe, the magnetization is still smaller than the expected value for saturated Dy$^{3+}$ (10.0 $\mu_{B}$). It is possible that the saturated moment is reduced due to CEF effect\cite{DyHo} or that there is another metamagnetic transition at fields higher than 90 kOe. An alternative explanation may be related to the deviation of local moments from \textit{c}-axis at low temperature as reported by a neutron study\cite{DyHo}. However, it is not obvious that the reported deviation would still exist under applied fields of up to 70 kOe. Magnetoresistance with \textit{H}$\parallel$\textit{c} shows highly non-monotonic behavior that is consistent with magnetization isotherm data. First, it increases rapidly at the first critical field observed in magnetization isotherm measurement and then decreases even more sharply at the second transition field. Finally, the resistivity stays constant at higher fields. Metamagnetic transition critical fields in the present work are close to those reported in a previous study on polycrystalline samples, although the features found here are much sharper\cite{TbDy}.

\subsection{HoCo$_{2}$Ge$_{2}$}

\begin{figure}[!ht]
\includegraphics[width=130mm, height=100mm]{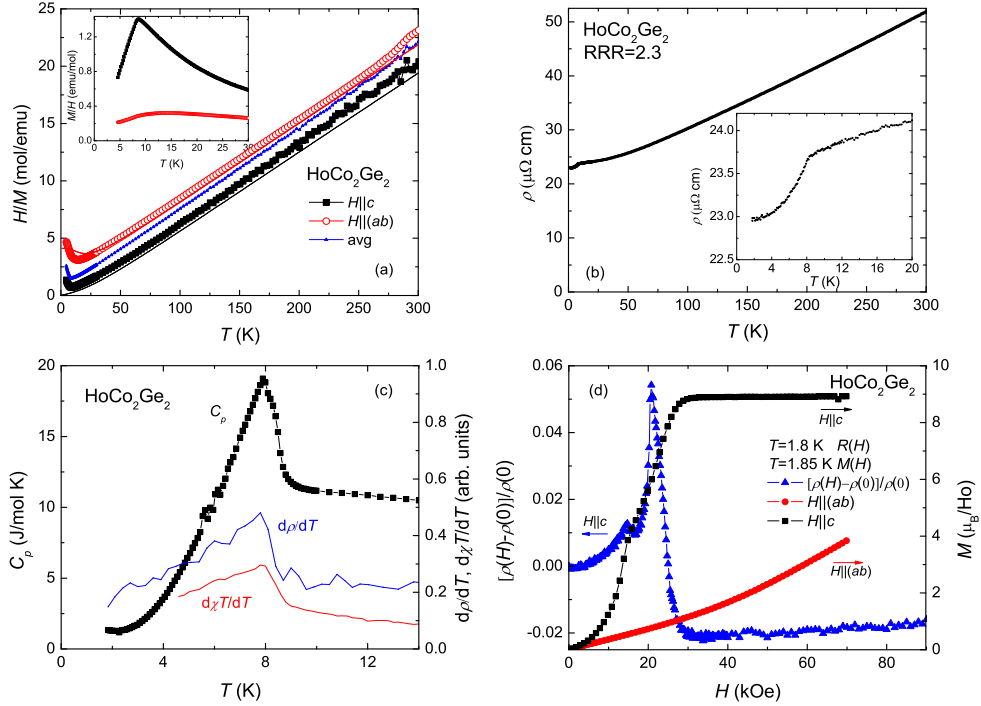}

\caption{Measurements of HoCo$_{2}$Ge$_{2}$ (a) Anisotropic and polycrystalline averaged inverse magnetic susceptibility measured at 1 kOe. Solid curves represent magnetic susceptibilities calculated from model based on CEF parameter. (inset: low-temperature magnetic susceptibility) (b) Zero-field resistivity (inset: low-temperature resistivity) (c) Zero-field specific heat on the left scale, d$\chi$\textit{T}/d\textit{T} and d$\rho$/d\textit{T} in arbitrary units. (d) Magnetization isotherms and normalized magnetoresistance.}
\label{Ho}
\end{figure}

Measurements made on HoCo$_{2}$Ge$_{2}$ are summarized in Fig.~\ref{Ho}. It has a smaller magnetic anisotropy than TbCo$_{2}$Ge$_{2}$ or DyCo$_{2}$Ge$_{2}$, yet $\chi_{c} > \chi_{ab}$ still holds. Linear fits of the inverse susceptibilities gives: $\Theta_{(ab)}$ = -21 K, $\Theta_{c}$ = 13 K and $\Theta_{ave}$ = -6.4 K. The effective moment is 10.6 $\mu_{B}$, which is the same as the expected value of 10.6 $\mu_{B}$ for Ho$^{3+}$ ion. A single transition at 7.8 K was observed. 

The temperature-dependent resistivity (Fig.~\ref{Ho}(b)) decreases in the vicinity of the transition temperature, which is consistent with magnetic susceptibility measurement. The RRR is 2.3.

There are two features in zero-field specific heat measurement (Fig.~\ref{Ho}(c)). The first one at 8.4 K is manifested by a change of slope and the second one, at 7.9 K, is associated with the local maximum in $C_{p}$. The fact that there exists two closely spaced transitions is consistent with an early neutron study\cite{DyHo}, albeit the exact transition temperatures are different. As these two transitions are very close in temperature, they are not clearly discernible in the derivatives of temperature-dependent magnetization and resistivity. In Table.~\ref{all data}, the ordering temperature is inferred from specific heat measurement. The magnetic entropy suggests that there may be a pseudo-triplet ground state below the ordering temperature. 

Magnetization isotherms measured at 1.85 K are shown in Fig.~\ref{Ho}(d). Comparing these data with those of DyCo$_{2}$Ge$_{2}$, in HoCo$_{2}$Ge$_{2}$, for \textit{H}$\parallel$\textit{c}, the middle plateau is much narrower and for \textit{H}$\parallel$\textit{(ab)}, the magnetization shows a positive curvature instead of the negative curvature found for DyCo$_{2}$Ge$_{2}$. The metamagnetic transition fields for HoCo$_{2}$Ge$_{2}$ are also different: 15 kOe and 23 kOe. Similar to the case in DyCo$_{2}$Ge$_{2}$, the saturated magnetization of HoCo$_{2}$Ge$_{2}$ is lower than the free ion value for Ho$^{3+}$. Neutron studies\cite{PrHo,DyHo} suggest that the size of the Ho$^{3+}$ moment is probably reduced at low temperature. Therefore, according to the present measurement, the reduced saturation moment for Ho$^{3+}$ at 1.85 K is 8.9 $\mu_{B}$/Ho$^{3+}$. The magnetoresistance is cusp-like at both metamagnetic transitions and it stays fundamentally unchanged for applied fields along the \textit{c}-axis above 30 kOe.

\subsection{ErCo$_{2}$Ge$_{2}$}

\begin{figure}[!ht]
\includegraphics[width=130mm, height=100mm]{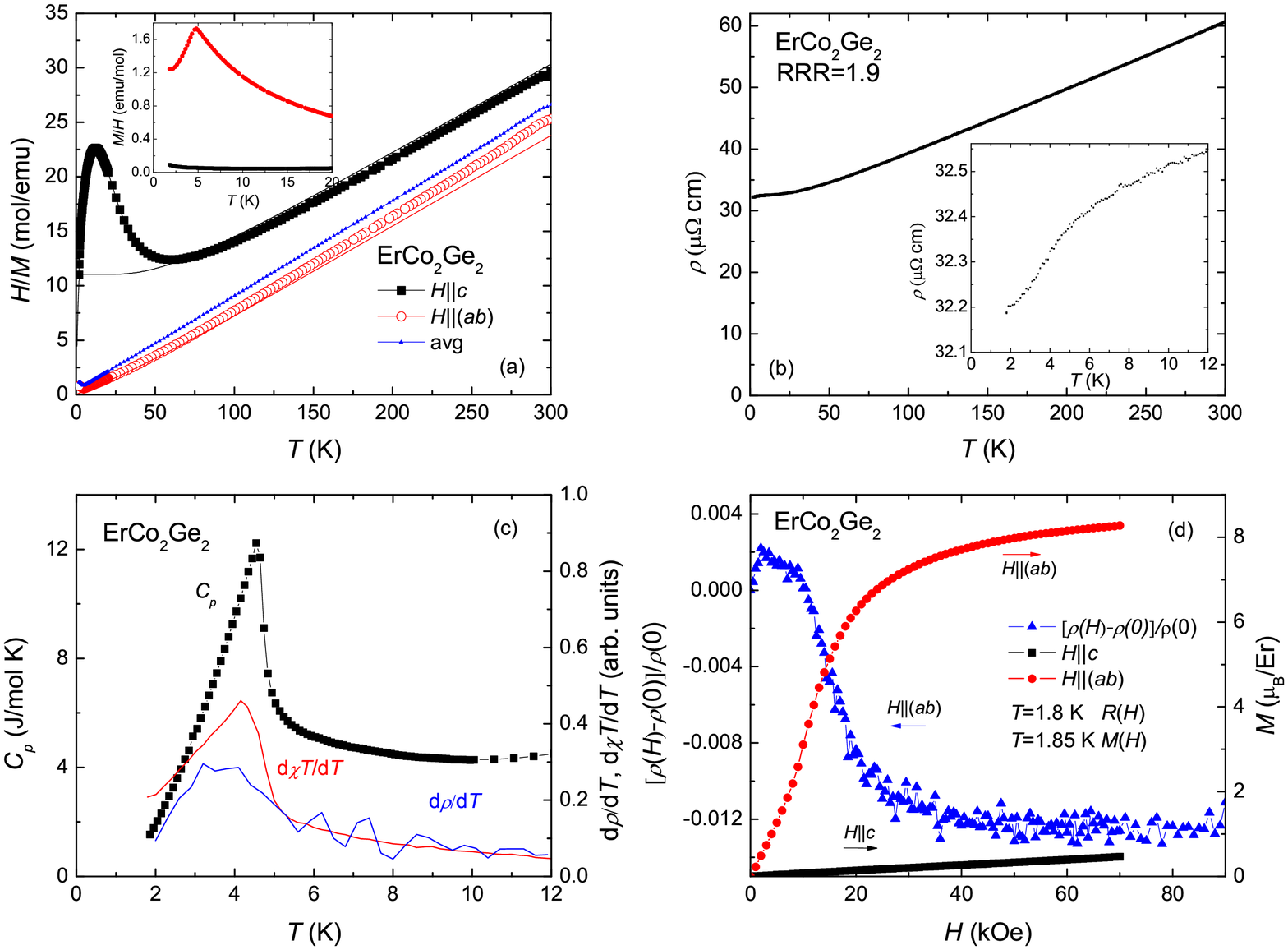}

\caption{Measurements of ErCo$_{2}$Ge$_{2}$ (a) Anisotropic and polycrystalline averaged inverse magnetic susceptibility measured at 1 kOe. Solid curves represent magnetic susceptibilities calculated from model based on CEF parameter. (inset: low-temperature magnetic susceptibility) (b) Zero-field resistivity (inset: low-temperature resistivity) (c) Zero-field specific heat on the left scale, d$\chi$\textit{T}/d\textit{T} and d$\rho$/d\textit{T} in arbitrary units. (d) Magnetization isotherms and normalized magnetoresistance.}
\label{Er}
\end{figure}

As shown in Fig.~\ref{Er}(a), in comparison to the previous local-moment-bearing rare earth members, the magnetic anisotropy is reversed in ErCo$_{2}$Ge$_{2}$ with $\chi_{ab}>\chi_{c}$. A high temperature linear fit on \textit{H}/\textit{M}(\textit{T}) gives $\Theta_{(ab)}$ = 14 K, $\Theta_{c}$ = -69 K and $\Theta_{ave}$ = -2.5 K. The effective magnetic moment yields 9.5 $\mu_{B}$, which is close to the theoretical value for trivalent Er (9.6 $\mu_{B}$). Transition temperature inferred from d$\chi\textit{T}$/d\textit{T} is 4.2 $\pm$ 0.6 K.

The temperature-dependent resistivity has a RRR of 1.9. It shows a weak, relatively smeared feature of the loss of spin-disorder scattering feature around the transition temperature. Nevertheless, in the plot of d$\rho$/d\textit{T} (Fig.~\ref{Er}(c)), the broad peak with a low signal-to-noise ratio makes it difficult to extract a clear signature of the transition temperature. 

The low temperature specific heat data exhibit a well-defined $\lambda$ shape, from which the transition temperature can be inferred to be 4.5 $\pm$ 0.1 K. Estimated magnetic entropy indicates a doublet below the ordering temperature, which is in agreement with an earlier study\cite{Ercp}.
 
In magnetization isotherm measurements, for \textit{H}$\parallel$\textit{c}, the magnetization increases linearly with applied field. When \textit{H}$\parallel$\textit{(ab)}, the magnetization tends to saturate above 60 kOe with a moment size being around 8 $\mu_{B}$/Er, which is smaller than its saturated value. In the lower field region, at around 9 kOe, the break in the slope of the in-plane magnetization increase seems to suggest a metamagnetic transition. This field corresponds to the observed change in slope of the magnetoresistance data. When conducting the experiment, special care was taken to ensure the direction of applied field is the same in magnetization isotherm and magnetoresistance measurement for the case of \textit{H}$\parallel$\textit{(ab)}. In this way, the effect of possible magnetic in-plane anisotropy is avoided.

\subsection{TmCo$_{2}$Ge$_{2}$}

\begin{figure}[!ht]
\includegraphics[width=130mm, height=100mm]{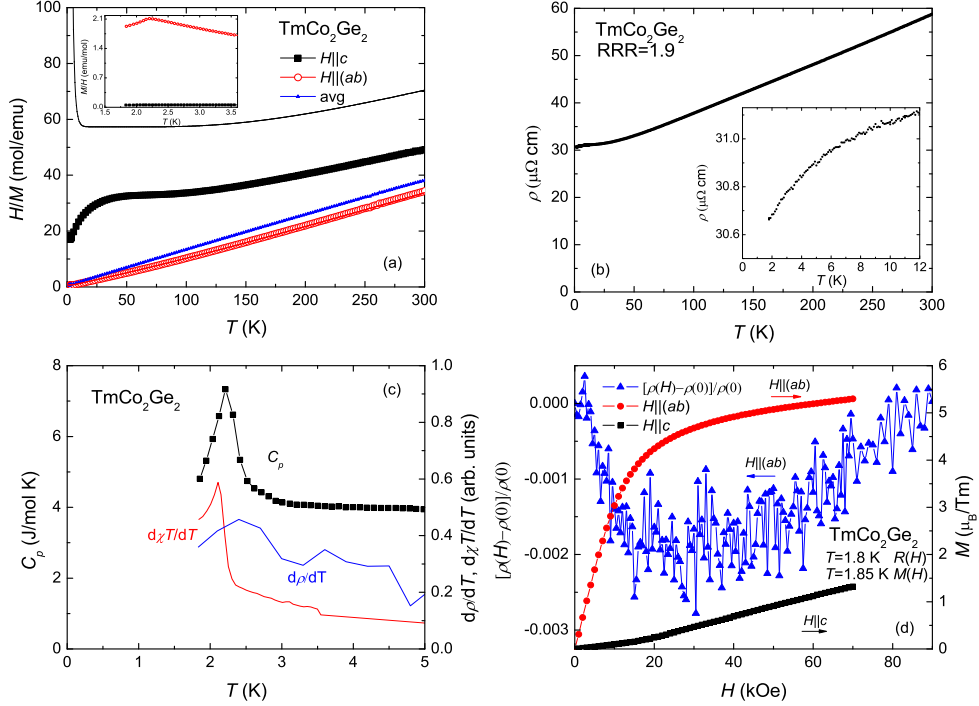}

\caption{Measurements of TmCo$_{2}$Ge$_{2}$ (a) Anisotropic and polycrystalline averaged inverse magnetic susceptibility measured at 1 kOe. Solid curves represent magnetic susceptibilities calculated from model based on CEF parameter. (inset: low-temperature magnetic susceptibility) (b) Zero-field resistivity (inset: low-temperature resistivity) (c) Zero-field specific heat on the left scale, d$\chi$\textit{T}/d\textit{T} and d$\rho$/d\textit{T} in arbitrary units. (d) Magnetization isotherms and normalized magnetoresistance.}
\label{Tm}
\end{figure}

The temperature-dependent magnetic susceptibility of TmCo$_{2}$Ge$_{2}$ manifests an even larger planar anisotropy than in the case of ErCo$_{2}$Ge$_{2}$ (Fig.~\ref{Tm}(a)). Just above T$_{N}$, $\chi_{(ab)}/\chi_{c} \sim 36$, which indicates an extreme planar anisotropy. A low ordering temperature of 2.2 K can be clearly seen. The effective moment calculated from a linear fit on high temperature polycrystalline inverse susceptibility yields 7.4 $\mu_{B}$, close to the theoretical value for Tm$^{3+}$ (7.6 $\mu_{B}$). The Curie-Weiss temperatures are: $\Theta_{(ab)}$ = 26 K, $\Theta_{c}$ = -220 K and $\Theta_{ave}$ = -2.3 K. Features observed here are consistent with antiferromagnetic ordering demonstrated by a recent neutron scattering work\cite{TmGondek}. However, their claim of Tm moments being parallel to \textit{c}-axis is clearly ruled out by our observations. It worth pointing out that their neutron diffraction pattern\cite{TmGondek} actually demonstrated that the Tm magnetic moments have at least a significant perpendicular component to the \textit{c}-axis with a well defined (001) Bragg peak emerging below the ordering temperature.  

Fig.~\ref{Tm}(b) shows the temperature-dependent resistivity data of TmCo$_{2}$Ge$_{2}$. Its derivative (shown in Fig.~\ref{Tm}(c)), similar to that of ErCo$_{2}$Ge$_{2}$, does not allow for an accurate determination of transition temperature. In Table~\ref{all data}, the transition temperatures of ErCo$_{2}$Ge$_{2}$ and TmCo$_{2}$Ge$_{2}$ are based on C$_{p}$(\textit{T}) and \textit{M}/\textit{H}(\textit{T}) measurements.

The zero-field specific heat shows a salient feature with transition temperature inferred to be 2.2 $\pm$ 0.1 K. Estimated magnetic entropy mostly comes from linearly extrapolated value in specific heat below 1.8 K, and the ground state may be a doublet.

Fig.~\ref{Tm}(d) shows the magnetization isotherms of TmCo$_{2}$Ge$_{2}$ measured at 1.85 K. For \textit{H}$\parallel$\textit{c}, the magnetization increases close to linearly with applied field below and above 20 kOe with slightly different slopes. For \textit{H}$\parallel$\textit{(ab)} there is a feature below 20 kOe that is most likely a metamagnetic transition, which is very broad due to T$_{N}$ being only slightly above the measurement temperature. Magnetoresistance, which was carefully measured at 1.8 K with the external field applied in the same direction as the in-plane \textit{M}(\textit{H}) measurement, although noisy, shows a very broad positive curvature. The magnetic field where magnetoresistance starts to increase coincides approximately with the critical field of the metamagnetic transition, observed in \textit{c}-direction. As mentioned above, the temperature of measurement is very close to the magnetic ordering temperature, possible metamagnetic transitions may be smeared by thermal excitation.

\section{Discussion}
In the RCo$_{2}$Ge$_{2}$ series, YCo$_{2}$Ge$_{2}$ and LaCo$_{2}$Ge$_{2}$ exhibit roughly temperature-independent magnetic susceptibility as expected for metals without local moment and magnetic ordering. The slight anisotropy of their Pauli/Landau-based magnetic susceptibility can be explained by the anisotropy in their Fermi surface. Since the electron count should be the same for these two members, the reversal of magnetic anisotropy is probably due to the difference in lattice parameter, which can end up changing the Fermi surface geometry. The electronic heat capacity for each is about 2-3 mJ/(mol-atomic) K, which is common for metals.

\begin{figure}[!ht]
\begin{center}
\includegraphics[width=92mm, height=75mm]{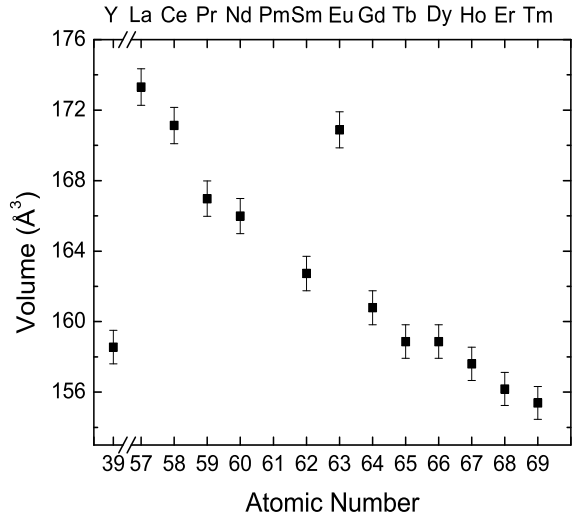}
\end{center}
\caption{Unit cell volumes of RCo$_2$Ge$_2$ (R = Y, La-Nd, Sm-Tm).}
\label{lattice}
\end{figure}

From the perspective of structure, lanthanide contraction is expected when substituting one rare-earth ion for another in an isostructural series of compounds. Lattice parameters refined from powder x-ray diffraction are shown in Table~\ref{table lattice parameter} and unit cell volumes are shown in Fig.~\ref{lattice}. There is a monotonic, almost linear contraction of unit cell volume with increasing atomic number across the trivalent lanthanides. EuCo$_2$Ge$_2$ exhibits a clear deviation from this trend, which is consistent with Eu being divalent instead of trivalent. As is often the case, unit cell volume of YCo$_2$Ge$_2$ is close to those of DyCo$_2$Ge$_2$ and HoCo$_2$Ge$_2$.

\begin{table}[!h]
\begin{center}
\begin{tabular}{|c|c|c|c|c|}
\hline
Compound & a (\AA) & c (\AA) & Volume (\AA$^3$) \\
\hline
YCo$_2$Ge$_2$&3.97&10.06&158.55\\
\hline
LaCo$_2$Ge$_2$&4.11&10.26&172.31\\
\hline
CeCo$_2$Ge$_2$&4.09&10.23&171.13\\
\hline
PrCo$_2$Ge$_2$&4.05&10.18&166.98\\
\hline
NdCo$_2$Ge$_2$&4.04&10.17&165.99\\
\hline
SmCo$_2$Ge$_2$&4.01&10.12&162.73\\
\hline
EuCo$_2$Ge$_2$&4.04&10.47&170.89\\
\hline
GdCo$_2$Ge$_2$&3.99&10.10&160.79\\
\hline
TbCo$_2$Ge$_2$&3.97&10.08&158.87\\
\hline
DyCo$_2$Ge$_2$&3.97&10.08&158.87\\
\hline
HoCo$_2$Ge$_2$&3.96&10.05&157.60\\
\hline
ErCo$_2$Ge$_2$&3.95&10.01&156.18\\
\hline
TmCo$_2$Ge$_2$&3.94&10.01&155.39\\
\hline

\end{tabular}
\end{center}
\caption{Lattice parameters and unit cell volumes of RCo$_{2}$Ge$_{2}$. The uncertainty is about 0.2\% for lattice parameter value.}
\label{table lattice parameter}
\end{table}

\begin{figure}[!ht]
\includegraphics[width=130mm, height=50mm]{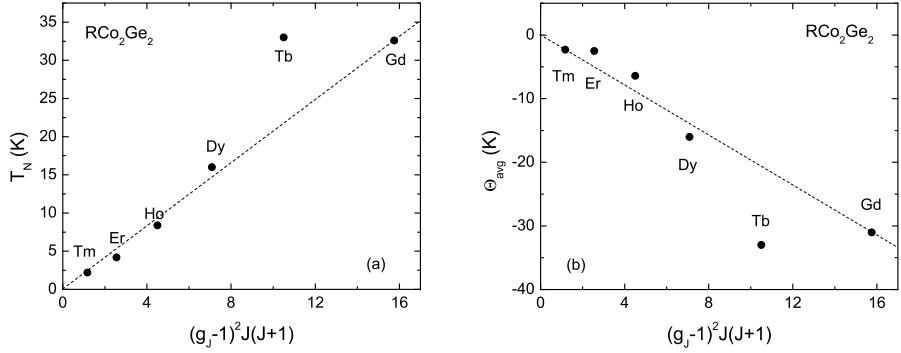}

\caption{Changes of (a) magnetic ordering temperature and (b) polycrystalline paramagnetic Curie-Weiss temperature with de Gennes parameter dG. The dashed lines represent simple de Gennes scaling normalized to the GdCo$_{2}$Ge$_{2}$ values.}
\label{dg}
\end{figure}

In discussing magnetic properties, within the mean field theory, both polycrystalline Curie-Weiss temperatures and magnetic ordering temperatures are predicted to be proportional to the de Gennes factor dG (dG = (g$_{J}$-1)$^{2}$J(J+1)) if the effects of CEF splitting of the Hund's rule ground state multiplet, J, are ignored\cite{book}. In practice, this proportionality is best seen for the heavy rare earth members (Gd-Tm). Fig.~\ref{dg} shows the obtained ordering temperature, T$_{N}$, and polycrystalline Curie-Weiss temperature, $\Theta_{ave}$, plotted against the dG factor. The ordering temperatures of heavy rare-earth members (Gd-Tm) do not follow the dG scaling. In particular, the ordering temperature of Tb is higher than, or at best comparable to, that of Gd, if
experimental uncertainties are included. In the plot of the Curie-Weiss temperature, Tb also deviates from the predicted trend. An early study on polycrystals\cite{mccall2} did show a rough de Gennes scaling in T$_{N}$ for members from Gd-Ho. The ordering temperatures inferred are close to those in the present study, except for GdCo$_2$Ge$_2$, which is much higher. Yet, in their study, they also reported a higher magnitude of Curie-Weiss temperature for Tb than Gd. A similar breakdown of de Gennes scaling had been investigated in the series of RRh$_{4}$B$_{4}$ compounds\cite{Noakes,Dunlap}. In that case, the peak of ordering temperatures in the de Gennes plot occurs at Dy instead of Gd. Theory had shown that with the CEF effect being taken into account, the ordering temperature can be enhanced to different extents according to the rare earth ion's total angular momentum. When a strong easy \textit{c}-axis anisotropy is present, Tb may become the new maximum in ordering temperatures. The deviation of $\Theta_{avg}$ from de Gennes scaling can be associated with the CEF effect as well. A breakdown of de Gennes scaling in a recent study on RNi$_{1-x}$Bi$_{2+y}$\cite{xiao} was also observed. However, it is still unclear why this breakdown is not a general observation for other rare earth intermetallic series with the same point symmetry and comparable single ion anisotropies\cite{RNi,RFe}.   

\begin{table}[!h]
\caption{Anisotropic Curie-Weiss temperatures, effective magnetic moment in paramagnetic state, magnetic transition temperatures and experimental value of B$_{2}^{0}$ of RCo$_{2}$Ge$_{2}$ (R = Y, La-Nd, Sm-Tm). Magnetic transition temperatures are shown for each measurement: T$_{M}$ is inferred from d$\chi\textit{T}$/d\textit{T}; T$_{\rho}$ is inferred from d$\rho$/d\textit{T}; T$_{C_{p}}$ is inferred from zero-field specific heat measurements and T$_{m}$ covers the range inferred by all three measurements. The highest magnetic transition temperature is the N\'{e}el temperature T$_{N}$. }
\begin{tabular}{p{0.8cm} p{0.8cm} p{0.8cm} p{0.8cm} p{0.8cm} p{1.2cm} p{1.2cm} p{1.2cm} p{1.2cm} p{1.1cm}}
\hline
\hline
R& $\Theta_{c}$ (K)& $\Theta_{(ab)}$ (K)& $\Theta_{avg}$ (K)& $\mu_{eff}$ ($\mu_{B}$)& T$_{M}$ (K)& T$_{\rho}$ (K)& T$_{C_{p}}$ (K)& T$_{m}$ (K)& B$_{2}^{0}$ (K)\\
\hline
Y& -& -& -& -& -& -& -& -& -\\
La& -& -& -& -& -& -& -& -& -\\
Ce& -& -& -& 2.6&& -& -&  $< 1.8$& -\\
Pr& 38& -180& -23& 3.5& 27.0(0.3)& 26.4(0.6)& 26.7(0.2) &26.7(0.9)& -9.4\\
Nd& 11& -50& -19& 3.6& 28.0(0.7)*& 27.7(0.7)& 27.9(0.2)& 27.9(0.9);  & -2.1\\
Sm& -& -& -& -& 17.0(0.8)& 16.7(1.2)& 16.5(0.2)& 16.7(1.2)& -\\
Eu& 7& 11& 10& 7.7& 22.2(0.3)& 22.1(0.3)& 22.1(0.2)& 22.1(0.4)& -\\
Gd& -36& -31& -32& 8.1& 33.2(1.8)& 31.7(1.5)& 33.1(0.2)& 32.6(2.4)& -\\
Tb& 37& -160& -33& 9.9& 33.0(1.0)& 32.8(0.9)& 33.3(0.1)& 33.0(1.1);& -4.0\\
 & & & & & 29.0(0.5)& 29.2(0.3)& 29.7(0.1)& 29.7(0.1); & \\
 & & & & & -& -& 28.8(0.1)&  28.8(0.1); & \\
 & & & & & 2.3(0.1)& -& 2.4(0.1)&  2.4(0.2) & \\
Dy& 28& -80& -16& 10.4& 16.0(0.7)& 15.9(0.5)& 16.5(0.2)& 16.0(0.7); & -1.4\\
 & & & & & 14.6(0.2)& 14.2(0.2) & 14.3(0.2)& 14.4(0.4) & \\
Ho& 13& -21& -6.4& 10.6& -& -& 8.4(0.1)& 8.4(0.1)& -0.4\\
 & & & & & 7.8(0.6)& 7.8(0.3)& 7.9(0.1)& 7.8(0.6); & \\
Er& -69& 14& -2.5& 9.5& 4.2(0.6)& -& 4.5(0.1)& 4.2(0.6)& 1.0\\
Tm& -220& 26& -2.3& 7.4& 2.2(0.2)& -& 2.2(0.1)& 2.2(0.2)& 5.0\\
\hline
\end{tabular}
*: There also exists a feature at 8.6(2.9) K that may be associated with magnetic transition.
\label{all data}
\end{table}

Except for Eu and Gd, all local moment bearing rare earth members in RCo$_{2}$Ge$_{2}$ exhibit magnetic anisotropy in their high temperature paramagnetic state. The CEF effect is identified as the leading reason for this anisotropy. Knowing that the rare earth ions in RCo$_{2}$Ge$_{2}$ possess a tetragonal point symmetry, if x, y and z-axis are chosen along conventional a, b and c-axis of the tetragonal structure, the CEF Hamiltonian can be expressed as follows\cite{CEFtet}:

\begin{equation}
H_{CEF} = B_{2}^{0}O_{2}^{0} + B_{4}^{0}O_{4}^{0} + B_{4}^{4}O_{4}^{4} + B_{6}^{0}O_{6}^{0} + B_{6}^{4}O_{6}^{4}
\end{equation}

\noindent
where $B_{n}^{m} = A_{n}^{m}\langle r^{n}\rangle\theta_{n}$ are the crystal field parameters and O$_{n}^{m}$ are the Stevens operator equivalents. A$_{n}^{m}$ are constants that reflect the strength of the CEF originating from the ions surrounding the central R ion. Although A$_{n}^{m}$ may change slightly when progressing across the rare-earth series, due to lanthanide contraction, generally, it can be viewed as a constant for different rare-earth elements, as it will not affect the trend significantly. $\langle$r$^{n}\rangle$ is the expectation value of r$^{n}$ for 4f electrons. $\theta_{n}$ are multiplicative factors which only need to be calculated once: $\theta_{2} = \alpha_{J}$; $\theta_{4} = \beta_{J}$ and $\theta_{6} = \gamma_{J}$. Previous studies\cite{wang,Boutron} showed that in a tetragonal symmetry, B$_{2}^{0}$O$_{2}^{0}$ is the leading term characterizing crystal electric potential energy and consequently the anisotropic magnetic behavior. Experimentally, B$_{2}^{0}$ can be determined from the high temperature anisotropic paramagnetic Curie-Weiss temperatures\cite{wang} by:

\begin{equation}
B_{2}^{0} = \frac{10}{3(2J-1)(2J+3)}k_{B}(\Theta_{(ab)} - \Theta_{c})
\end{equation}

\noindent
where J is the total angular momentum of Hund's rule ground state for the specific rare-earth ion under study. This equation provides a way to determine B$_{2}^{0}$ from the data listed in Table~\ref{all data}. The sign of B$_{2}^{0}$ reflects whether the compound has an easy-axis or easy-plane anisotropy. In addition, as described before, B$_{2}^{0}$ can also be theoretically calculated through:

\begin{equation}
B_{2}^{0} = \langle r^{2}\rangle A_{2}^{0}\alpha_{J}
\end{equation}

\begin{figure}[!ht]
\begin{center}
\includegraphics[width=75mm, height=61mm]{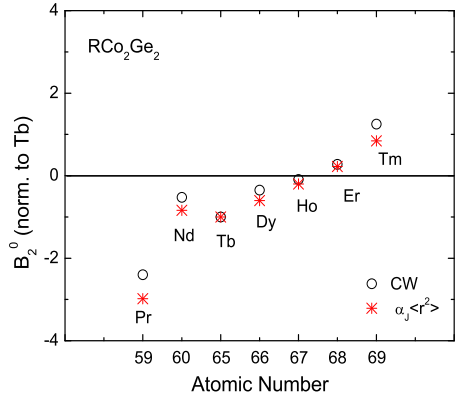}
\end{center}
\caption{Normalized (both experimental and theoretical values for TbCo$_{2}$Ge$_{2}$ are -1) CEF parameters B$_{2}^{0}$. }
\label{CEF}
\end{figure}

If A$_{2}^{0}$ is viewed as a constant, B$_{2}^{0}$ can be calculated by adopting theoretical values of $\langle$r$^{2}\rangle$ \cite{Freeman}and $\alpha_{J}$\cite{Hutchings}. As $\langle$r$^{2}\rangle$ is definite a positive number and $\alpha_{J}$ is theoretically calculated, the easy magnetization direction still depends on the sign of $A_{2}^{0}$, which can lead to different anisotropic properties\cite{RNi, RFe, RAgSb2, TbC}. Fig.~\ref{CEF} shows normalized experimental and theoretical values of B$_{2}^{0}$. Normalization was done to ensure both theoretical and experimental value of B$_{2}^{0}$ for Tb$^{3+}$ are -1, which can help to factor out the influence of A$_{2}^{0}$ not being taken into consideration. Fig.~\ref{CEF} plots the $B_{2}^{0}$ values inferred from our anisotropic values using Eqn.(2) as well as those calculated using Eqn.(3). Two data sets agree quite well which indicates that the anisotropy in this series compounds is indeed governed by the leading term of the CEF effect.

In order to see how much of the $T > T_{N}$ magnetic anisotropy the $B_{2}^{0}$ term captures for these materials, we can look at a very simplified Hamiltonian given by

\begin{equation}
H = B_{2}^{0}O_{2}^{0} + g_{J}\mu_{B}JH
\end{equation}

\noindent
where the second term is the Zeeman energy. Using this, we have calculated the paramagnetic susceptibility that would be obtained from such a Hamiltonian (see, for example work done by P. Boutron\cite{Boutron}), where B$_{2}^{0}$ determined from experimental Curie-Weiss temperatures is used. In Fig.4, 5 and 9-13, calculated magnetization curves were shown for RCo$_{2}$Ge$_{2}$ (R = Pr, Nd, Tb-Tm) in solid curves. As a first approximation, these curves capture the basic anisotropies and temperature-dependence. The discrepancies demonstrate that the influence of higher order crystal field parameters as well as RKKY interaction need to be considered so as to provide a more accurate physical interpretation for magnetic anisotropy, especially for members like PrCo$_{2}$Ge$_{2}$ and TmCo$_{2}$Ge$_{2}$. Quantitatively, magnetic measurements on diluted magnetic rare earth ions in either YCo$_{2}$Ge$_{2}$ or LaCo$_{2}$Ge$_{2}$ would be needed to provide more detailed information about the CEF splitting and, as a result, provide a better understanding of the effects of the exchange interaction. 

Comparing the RT$_{2}$Ge$_{2}$ (T = Fe, Co, Ni) series, for non-magnetic members, the electronic specific heats of YCo$_{2}$Ge$_{2}$ and LaCo$_{2}$Ge$_{2}$ are close to those obtained in RNi$_{2}$Ge$_{2}$\cite{RNi}. Furthermore, as the magnetic susceptibilities of RCo$_{2}$Ge$_{2}$ (R = Y and La) are also close to their counterparts in Ni series, the density of states at the Fermi energy is likely to be similar for RCo$_{2}$Ge$_{2}$ and RNi$_{2}$Ge$_{2}$. On the other hand, in RFe$_{2}$Ge$_{2}$\cite{RFe}, the non-magnetic $\gamma$ is nearly six times larger, which may be evidenced by a magnetic fluctuation on the Fe site\cite{LuFe}. All three series exhibit generally the same magnetic anisotropy where the paramagnetic easy magnetization direction switch from axial (R = Pr, Nd, Tb-Ho) to planar (R = Er, Tm) following a positive A$_{2}^{0}$. The ordering temperatures of RT$_{2}$Ge$_{2}$ (T = Fe, Co, Ni) for heavy rare earth members shows a rough trend of \textit{T}$_{Co} >$ \textit{T}$_{Ni} >$ \textit{T}$_{Fe}$\cite{RNi,RFe,TbNi}. Given the enhanced magnetic susceptibility of RFe$_{2}$Ge$_{2}$ compared with the other two series, it is interesting to have, in fact, lower ordering temperature for RFe$_{2}$Ge$_{2}$. All of CeT$_{2}$Ge$_{2}$ (T= Fe, Co and Ni) show Kondo lattice behavior with moderately enhanced $\gamma$ values\cite{RNi,CeFe,CeNi}. Among these three compounds, CeCo$_{2}$Ge$_{2}$ has the highest Kondo temperature. Detailed band structure calculation may be helpful in illuminating the changes in these RT$_{2}$Ge$_{2}$ series as T is varied from Fe to Co to Ni.

\section{Conclusion}

In this work, single crystalline ternary compounds RCo$_{2}$Ge$_{2}$ (R = Y, La-Nd, Sm-Tm)  were grown using a self-flux method and characterized by x-ray powder diffraction, temperature- and field-dependent magnetization, temperature- and field-dependent resistivity and zero-field specific heat measurements. Magnetic ordering temperatures were determined down to 1.8 K by d$\chi$\textit{T}/d\textit{T}, d$\rho$/d\textit{T} and zero-field specific heat. All local-moment bearing rare-earth members, except for Ce, order antiferromagnetically above 2 K with the highest T$_{N}$ value being 33.0 K (Tb) and the lowest being 2.2 K (Tm). YCo$_{2}$Ge$_{2}$ and LaCo$_{2}$Ge$_{2}$ are Pauli paramagnets. Ce is trivalent at high temperature in CeCo$_2$Ge$_2$. Although it does not appear to order magnetically, it does show two weak anomalies at 15 K and 2.3 K, of which we are uncertain about the origin. An enhanced $\gamma$ value, together with other transport measurements, is consistent with a Kondo screening effect with $\gamma \sim$ 90-103 mJ/mol K$^{2}$ and T$_{K} \sim$ 100 K. More than one magnetic phase transition was observed in RCo$_{2}$Ge$_{2}$ (R = Tb-Ho). Magnetic anisotropies were observed for all members, except for Gd$^{3+}$ and Eu$^{2+}$ with half-filled 4f shells (\textit{L} = 0). This anisotropy can be explained quite well by CEF theory, where crystal field parameter B$_{2}^{0}$ is the leading term deciding the easy direction and the size of anisotropy. The de Gennes scaling does not hold for heavy rare-earth members, which may partially be explained by a strong CEF effect. 

\section*{Acknowledgement}

We would like to thank W. Jayasekara, H. Hodovanets, A. Thaler and Greg Dyer for useful discussions and experimental assistances. We would also like to thank A. Kreyssig for not only providing general discussion, but also providing key understanding about the existing TmCo$_{2}$Ge$_{2}$ scattering data and analysis. Work done at Ames Laboratory was supported by US Department of Energy, Basic Energy Sciences, Division of Materials Sciences and Engineering under Contract NO. DE-AC02-07CH111358. V.T. and X.L. would like to acknowledge support from AFOSR-MURI Grant No. FA9550-09-1-0603. 

\bibliographystyle{elsarticle-num}

\end{document}